\documentclass[acmsmall]{acmart}

\usepackage{hyperref}
\usepackage{booktabs} 
\usepackage[ruled,linesnumbered]{algorithm2e}
\usepackage{url}
\usepackage{xcolor}
\usepackage{subfig}

\usepackage{amsthm}
\usepackage{amsmath}
\usepackage{enumitem}
\usepackage{xspace}
\usepackage{booktabs}
\usepackage{multirow}
\usepackage{subfig}
\usepackage{tablefootnote}
\usepackage{balance}
\usepackage{bibunits}

\newcommand\ab[1]{\textcolor{red}{Aline: #1}}
\newcommand\tds[1]{\textcolor{black}{#1}}

\newcommand\hide[1]{}
\newcommand{\tipo}{\textit{PODS}\xspace}
\newcommand{\trend}{data-trend\xspace}
\newcommand{\trends}{data-trends\xspace}
\newcommand{\Trend}{Data-trend\xspace}
\newcommand{\Trends}{Data-trends\xspace}
\newcommand\todo[1]{\textcolor{olive}{#1}}
\renewcommand\todo[1]{}

\newcommand\challenge[1]{(\textit{Challenge~#1})\xspace}

\newcommand\lineref[1]{\emph{line #1}\xspace}

\renewcommand{\paragraph}[1]{\vspace{0.1cm}\noindent \textbf{#1}}

\theoremstyle{definition}
\newtheorem{define}{Definition}

{\begin{list}{$\bullet$}{%
	\setlength{\labelsep}{2pt}\setlength{\leftmargin}{0pt}%
	\setlength{\labelwidth}{0pt}%
	\setlength{\listparindent}{0pt}}}
{\end{list}}

{\begin{list}{\noindent}{%
	\setlength{\leftmargin}{0pt}%
	\setlength{\labelwidth}{0pt}%
	\setlength{\listparindent}{0pt}}}
{\end{list}}

\hypersetup{draft}

\def\BibTeX{{\rm B\kern-.05em{\sc i\kern-.025em b}\kern-.08emT\kern-.1667em\lower.7ex\hbox{E}\kern-.125emX}}
    
\setcopyright{acmcopyright}

\begin{document}

%
\title{
\tds{Effective} Discovery of Meaningful Outlier Relationships}

%

\author{Aline Bessa}
\email{aline.bessa@nyu.edu}
\author{Juliana Freire}
\email{juliana.freire@nyu.edu}
\affiliation{%
  \institution{New York University}
  \streetaddress{2 Metrotech Pl, 10th floor}
  \city{Brooklyn}
  \state{New York}
  \postcode{11201}
}

\author{Tamraparni Dasu}
\email{tamr@research.att.com}
\author{Divesh Srivastava}
\email{divesh@research.att.com}
\affiliation{%
  \institution{AT\&T Labs-Research}
  \streetaddress{1 AT\&T Way}
  \city{Bedminster}
  \state{New Jersey}
  \postcode{07921}
}

%

\begin{abstract}
%

We propose \tipo (\textit{Predictable Outliers in Data-trendS}), a
method that, given a collection of temporal data sets, derives data-driven
explanations for outliers 
by identifying \textit{meaningful} relationships between them. 
%
First, we formalize the 
notion of meaningfulness, which so far has been informally framed in
terms of explainability.
Next, since outliers are rare and it is difficult to determine whether their relationships are meaningful, we develop a
new criterion that does so by checking if these relationships could have
been predicted from non-outliers, i.e., if \textit{we could see the
outlier relationships coming}. 
%
Finally, searching for meaningful outlier
relationships between every pair of data sets in a large data collection is  computationally
infeasible. To address that, we propose
%
an indexing strategy that  prunes irrelevant comparisons
across data sets, 
making the approach scalable. 
We present the results of an experimental evaluation using real data sets
and different baselines, which demonstrates the effectiveness,
robustness, and scalability of our approach.
\end{abstract}

\ccsdesc[500]{Information systems~Data analytics}
\ccsdesc[500]{Information systems~Data cleaning}
\ccsdesc[300]{Mathematics of computing~Exploratory data analysis}

%
\keywords{outlier explanation, urban data analysis, data sets, data cleaning}

%

%
\maketitle

\section{Introduction}
\label{sec:intro}
As data volumes continue to grow, data-driven policies and decisions
are becoming the norm. 
Consider, for instance, urban data.  
Increasingly, cities around the world are collecting and publishing open  data sets
that capture different aspects of urban
environments~\cite{barbosa@bigdata2014, opendata@book2013,
  chicagoopendata, nycopendata, sfoopendata, parisopendata,
  rioopendata}. 
By exploring these data sets and their relationships,
it is possible to better understand how different urban components interact.
This, in turn, can inform policies, make cities more efficient, and
improve their residents' lives. 
A notable example of data-driven policy making is the New York City
(NYC) Vision Zero initiative~\cite{vision-zero}, which aims to make
streets safer. Based on the observation that there is a relationship
between \textit{high} traffic speed and \textit{large} numbers of both
traffic accidents and pedestrian fatalities, one intervention
implemented by the City government was to reduce the speed limit on
the streets.
Many other actions have been informed by data, including changes to
policing strategies in areas with high foreclosure
rates~\cite{ellen2013foreclosures} and government spending on
subsidized housing~\cite{schill2002revitalizing}.

Hypotheses that serve as the basis for a decision can be tested by
looking for corresponding relationships in the underlying data. But
when a large number of data sets is available, we also have the
opportunity to derive new hypotheses by uncovering previously unknown
relationships.
The Data Polygamy approach~\cite{chirigati@sigmod2016} took a first
step in this direction, and proposed a new technique to identify
spatio-temporal relationships among data sets. 
This approach, however, has a limitation: it does not discover relevant 
relationships involving outliers. These relationships are important
because they capture events which are infrequent but have
strong effects, such as ``uptick in foreclosures leads to increase in
crime''~\cite{ellen2013foreclosures}.
%

The identification of outlier-based relationships is also useful in
the context of data cleaning. 
Typically in data analysis and predictive modeling, outliers are
removed or dampened in order to become acceptable
values~\cite{dasu2003}.  However, when outliers are not data errors
and carry relevant semantic information, their removal can lead to
unreliable results or statistical distortions~\cite{dasu@kdd2014}. 
%
Determining whether outliers should be cleaned requires domain
knowledge.  However, even when such knowledge is available for large data collections,  
or for complex analyses that involve multiple data sets with varying
levels of quality, manually inspecting the outliers is often
infeasible.
The presence of a relationship between outliers can not only alert an
expert about a previously-unknown interaction, leading to
new insights, but may also provide a potential explanation for the
outliers. This, in turn, can help guide the decision as to whether the outliers should be
\textit{cleaned}. 

In this paper, we define and study the problem of detection of
meaningful outlier relationships for temporal data.

\begin{figure}[t]
\centering
\includegraphics[width=0.7\columnwidth]{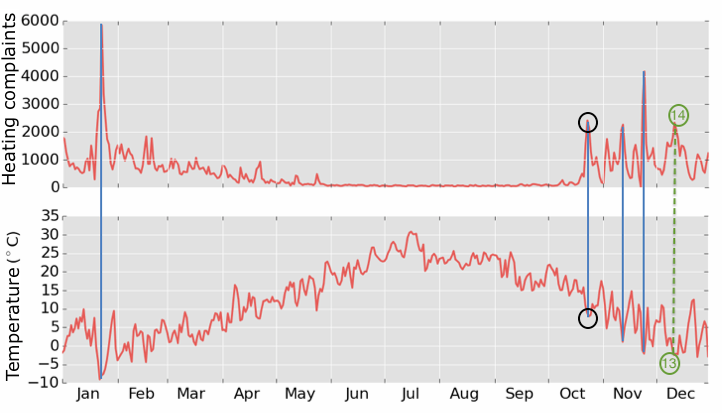}
\caption{Daily values for temperature and heating complaints in NYC in 2013. The number of heating complaints peaked over 2,000 on five
  different days, roughly when large drops in temperature were observed. The four solid blue lines connect extreme heating complaint values to their co-occurring temperatures, 
and the dashed green line connects a drop in temperature (December, 13) to a slightly mismatched peak in complaints (December, 14). These two dates are indicated in the  
 green circles. The black circles correspond to the first peak in complaints in October and its co-occurring temperature. }
\label{subfig:intro-raw}
\end{figure}

\paragraph{Outlier Relationship Discovery: Challenges} There are many
challenges involved in the discovery of meaningful outlier
relationships in temporal data sets. 
To devise a method that automatically detects relationships, we need a
formal and computable definition of \textit{meaningfulness} that
reflects (or that is close) to users' intuition
\challenge{1}.

%
Another challenge comes from the fact that \emph{aligned} outliers are rare.
Consider the plots in Figure~\ref{subfig:intro-raw}, which contrast
the number of daily New York City (NYC) 311 heating complaints, which
indicate lack of heat or hot water in residential
buildings~\cite{311-heating-complaint}, with average temperatures.
\hide{These complaints reflect inadequate living conditions that should be
addressed by the City.}  
Note that  
there is a \textit{relationship
between outliers in the 311 heating complaints and
temperature}: 
peaks in complaints often correspond to abrupt drops in
temperature. This suggests that these outliers correspond to
real-world events that can be \emph{explained}, and thus are unlikely to be data
errors.  
%
%
Since any given data set contains, by definition, a
small number of outliers, there are even fewer aligned outliers
between two data sets. 
Thus, it is difficult to determine if outlier alignments are related
in a \textit{meaningful} way, or if their interaction is simply
coincidental \challenge{2}.

Correlation metrics such as Pearson or Rank correlation coefficients
are widely used to identify relationships between different indicators. However, they are not
reliable for small sample sizes~\cite{CRR73}. Consequently, these
metrics cannot be applied in the identification of relationships
involving outliers, as they consist of few data points.
On the other hand, if we compute Pearson or Rank correlations over all data points, the discovered correlations (if any) may not correspond to the
relationship between the aligned outliers. 
Consider again the plots in Figure~\ref{subfig:intro-raw}.
If we take into account only the days with abnormally-high
numbers of complaints, there is a strong relationship with aligned 
drops in temperature.
However, when all days are taken into account, the relationship
between these two variables is much weaker: while the number of
complaints remains almost constant over the warmer months, the
temperature keeps varying, attenuating the relationship that can be
observed across extreme values. 

In data that represents temporal processes, an event that is reflected
as an outlier may influence (or be influenced by) other events that
occur in nearby time intervals.
As shown in Figure~\ref{subfig:intro-raw}, while some large
temperature drops and complaint peaks co-occur on the same day (blue
solid lines), sometimes they \textit{occur in close temporal
  proximity} (green dashed line).  As another example, if it rains
heavily for a few days, effects of the abnormal rainfall can be felt
even after the rain subsides.
It is thus important to detect extreme values that may not be
perfectly aligned in time but are temporally close nonetheless,
otherwise relevant explanations may be missed \challenge{3}.

Last but not least,  finding meaningful outlier
alignments \textit{at scale} is also difficult \challenge{4}.  
In a collection of data sets, the number of possible pairwise
relationships between attributes across data sets is quadratic in the
total number of attributes. For large collections, assessing all pairs
to detect meaningful outlier relationships can then be prohibitively expensive.

\begin{figure}[t]
\centering
\includegraphics[width=0.6\columnwidth]{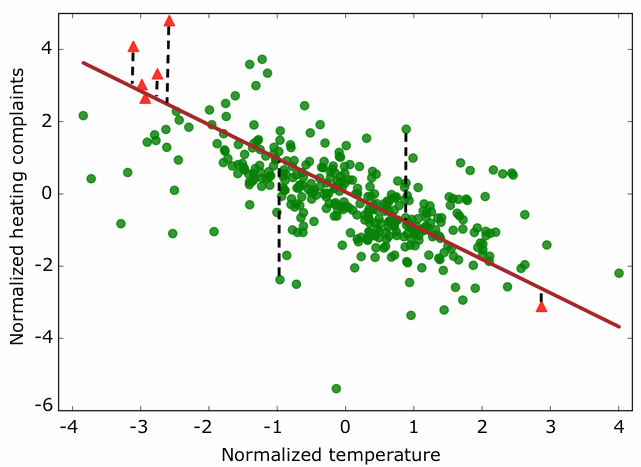}
\caption{Scatterplot of normalized values for heating complaints and temperature. Aligned outliers are represented as red triangles; the 
remaining aligned values, as green circles. 
The red line represents a
\trend that captures the relationship between increased number of complaints and large temperature drops.
Dashed black lines indicate estimation errors with respect to the \trend.
Aligned values that are geometrically close  do not need to be temporally near---the aligned outliers
 and inliers in the upper left quadrant, for example, belong to different months. 
}
\label{subfig:intro-scatterplot}
\end{figure}

%


\paragraph{Our Approach.}
To address \challenge{1}, we formalize the notion of
\textit{meaningfulness}. We say that an outlier relationship is
meaningful if it can be predicted from  nearby non-outliers, i.e., patterns
among outliers are coherent with those among
near-outliers. This corresponds to the intuition that looking at the
data, one can gradually see the relationship coming.
We verify this intuition through a user study described in
Section~\ref{setup-and-experiments}.




%
%

  
Given a collection of temporal data sets, we must first align them
temporally.  We \emph{introduce a scoring representation that takes into
account the cumulative effects of outliers}, thus enabling the
alignment of both co-occurring and temporally-close outliers
\challenge{3}.
Next, we assess the relationships between attributes that
contain aligned outliers. Given that there can be a large number of
pairs, we \emph{introduce an index that effectively prunes useless
  pairs}, thus making the approach scalable for large data set
collections \challenge{4}. 

%
For each pair of attributes that share aligned outliers, we
search for \textit{\trends} over \textit{the space of all aligned
  values} to assess if the outlier relationship is not
coincidental. 
While temporal trends, common in time series
studies~\cite{hamilton1994time}, capture the variation in values of
one attribute over time, a \textit{\trend} captures the
\textit{relationship between values of a pair of attributes}.
In fact, \tipo only uses timestamps to find alignments between values from 
 different data sets; time information is not used in the search for \trends. 
%
%
 To verify the existence of a \trend, we \emph{propose a new strategy
   based on weighted linear regression models}~\cite{CRR73} that
 ensures a better fit for aligned outliers and aligned near-outliers
 \challenge{2}.
 The intuition behind this fitting strategy is that real-world
 mechanisms and processes tend to be monotonic, and important
 relationships usually behave as trends. 
Consequently, patterns that are observed among aligned outliers are
likely to also materialize across aligned near-outliers.
Our strategy 
is illustrated by the red line in
Figure~\ref{subfig:intro-scatterplot}.
Finally, when a \trend is detected, we assess if it provides evidence
that the outlier alignments are statistically significant: we check
whether the \trend can be used to accurately predict the outliers. In
the case of Figure~\ref{subfig:intro-raw}, for example, the pattern between
temperature drops and heating complaint peaks is detected by \tipo as
statistically significant.

 
\paragraph{Contributions.} We propose \tipo (\textit{Predictable
  Outliers in Data-trendS}), a method that, given a collection of
temporal data sets, discovers meaningful outlier relationships across
them which can then be used to explain the outliers.
While the problem of explaining data and queries has been explored in
previous
work~\cite{bailis2017macrobase,chirigati@sigmod2016,roy@vldb2015,wu2013scorpion,zhang@edbt2017},
our approach is, to the best of our knowledge, the first to search for
explanations by discovering relevant outlier alignments, which are
generally rare and hard to evaluate statistically.

Our main contributions can be summarized as follows: 
\begin{itemize}
\item We propose a
formal definition for meaningful outlier relationships and we also
formulate the problem of detecting these relationships;
\item We design a method for detecting \trends that can be used to
  ascertain the statistical significance of outlier alignments,
  which can serve as a means to explain extreme events and data anomalies;
\item We introduce a representation strategy that takes
temporal cumulative effects into account, enabling the 
alignment of values that occur in different yet close time intervals 
and allowing the efficient detection of relationships for events that have
lingering effects;
\item We design an index that prunes attribute pairs that do not
share aligned outliers, thus making the approach scalable to
large collections of data sets;
\item We have carried out an extensive
experimental evaluation, contrasting \tipo with a variety of baselines over real urban data sets from 
New York City, and report results that demonstrate the effectiveness, scalability, and robustness 
of our approach. We also discuss the results of a user study which
supports our hypothesis that our definition of meaningfulness is
intuitive and corresponds to users' expectations. 
\end{itemize}

%

\paragraph{Outline.} The problem we address is formally defined in
Section~\ref{sec:problem-definition}. 
Our approach is described in Section~\ref{general-components}.
We report the setup and results of a comprehensive experimental evaluation in
Section~\ref{setup-and-experiments}.
Related work is discussed in Section~\ref{related-work}. 
We conclude in Section~\ref{conclusions} with a summary of our
findings and plans for future work.

\section{Definitions and Problem Statement}
\label{sec:problem-definition}

 
To the best of our knowledge, the problem of determining whether a relationship across aligned outliers from different sources of data is \textit{meaningful} 
has never been formally stated. 
In fact, the notion of meaningfulness has been framed so far in terms of explainability, i.e., \textit{an 
outlier relationship is meaningful if outliers from one data set \textit{explain} abnormal points in another data}~\cite{dasu@kdd2014,wu2013scorpion}. 
%
%
By formally defining meaningful outlier relationships, we restrict the scope of their detection while 
capturing their intuitive aspects, such as the connection with outlier explainability. 
%




In what follows, we introduce basic concepts required to formally state the problem. 
Let $D$ denote a data set, $X$ be a numerical attribute in $D$, and
$x^i$ be the value of attribute $X$ associated with a certain timestamp $t_i$. 

\begin{define}[\textbf{Outlier Detection
    Function}] \label{def:outlier-detection-function} An
  \textit{outlier detection function} $\psi_X$ maps attribute values
  $x^i$ to representation scores $u^i$ that reflect the degree of
  \emph{outlierness} of values $x^i$.
Function $\psi_X$ must satisfy the following requirement: higher (absolute) $u^i$ scores correspond to more severe outliers. 
\end{define}
\noindent Note that this requirement is flexible, allowing for different notions
of outlierness.  For example, it is possible to apply functions to
determine outliers based on context or global criteria.  Standardized
z-scores~\cite{CRR73}, 
i.e., the number of standard deviations value $x^i$ is from the mean, 
 and LoOP scores~\cite{kriegel@cikm2009} are
examples of suitable scores $u^i$. 

\begin{define}[\textbf{Outliers and Inliers}] \label{def:outlier-inlier}
Let  $\theta^{+}_X \in \mathbb{R}^+$ and $\theta^{-}_X \in \mathbb{R}^-$ be positive and negative \textit{outlier thresholds} for outlier detection 
function $\psi_X$. 
Score $u^i$ identifies an \textit{outlier} if $u^i > \theta^{+}_X$ or if $u^i < \theta^{-}_X$; otherwise, it identifies an \textit{inlier}. 
\end{define}
\noindent Suitable outlier thresholds depend on the semantics of function $\psi_X$ and the scores $u^i$ it derives.
\tds{If an attribute $X$ is normally distributed and represented with z-scores, for example, $\theta^{+}_X = -\theta^{-}_X = 3$ are commonly used as thresholds for outlier 
detection~\cite{CRR73}.
We use two outlier thresholds, $\theta^{-}_X$ and $\theta^{+}_X$,  because the distribution of $X$  may not be symmetric. Alternatively, the analysis may only concern outliers in
 one direction (i.e., either very high or very low values). In both cases, it is convenient to have different thresholds for positive and negative values.} 
%
%

%


\begin{define}[\textbf{Aligned Scores}] \label{def:aligned-scores}
Let $U_1$ and $U_2$ be the representations for attributes $X_1$ and $X_2$,
obtained by outlier detection functions $\psi_{X_1}$ and $\psi_{X_2}$ respectively. 
Given scores $u_1^i \in U_1$ and $u_2^j \in U_2$, $u_1^i$ and $u_2^j$  are aligned scores if they are associated with the same timestamp, 
i.e., $t_i = t_j$.  
\end{define}
\noindent \tds{In Figure~\ref{subfig:intro-raw}, for example,  each temperature value can be associated through a timestamp alignment with a number of heating complaints. 
 This alignment in time is preserved when the outlier functions are applied to the two attributes, resulting in corresponding scores that are also aligned in time. }

\begin{define}[\textbf{Aligned Outliers}] \label{def:aligned-outliers} 
Let $U_1$ and $U_2$ be the representations for attributes $X_1$ and $X_2$, and 
$O_1 \subset U_1$ and $O_2 \subset U_2$ be subsets comprising all
outliers of $U_1$ and $U_2$ respectively. 
Given scores $o_1^i \in O_1$ and $o_2^j \in O_2$, $o_1^i$ and $o_2^j$  are aligned outliers if they are associated with the same timestamp, 
i.e., $t_i = t_j$.  
\end{define} 
\noindent 
\tds{When the scores for both attributes happen to be outliers, the resulting aligned scores are said to be aligned outliers. 
As a concrete example, consider the data containing daily
values for temperature and heating complaints in Figure~\ref{subfig:intro-raw}. 
 We plot their aligned z-scores (days work as timestamps in this example) in Figure~\ref{subfig:intro-scatterplot}, 
and use a single outlier threshold $\theta_X = 2$. Each red triangle corresponds to aligned outliers 
$o_1^i$ and $o_2^j$ such that $|o_1^i| > 2$ and $|o_2^j| > 2$.}

\begin{define}[\textbf{Meaningful Outlier Relationships}] \label{def:meaningful} 
Let $U_1$ and $U_2$ be the representations for attributes $X_1$ and $X_2$. 
If there is a statistical model that adequately fits the aligned scores of $U_1$ and $U_2$, and also fits 
the aligned outliers of $O_1 \subset U_1$ and $O_2 \subset U_2$ in particular, we say that there is a \textit{meaningful 
outlier relationship} involving these outliers. 
\end{define}
\noindent In other words, meaningful outlier relationships are
\textit{predictable} given a model between $U_1$ and $U_2$ that
adequately fits the observations, i.e., \textit{we can see the aligned 
  outliers coming}.  Intuitively, an adequate model between $U_1$
and $U_2$ is evidence of meaningfulness because there is an 
expectation that the outlier alignments are not due to chance alone: they follow a pattern that is also present in closeby parts of the data.  

\paragraph{Problem Statement.} Given a collection of data sets
$\mathcal{D}$ and the union of their attributes $\{X_i, 1 \leq i \leq n\}$, outlier
detection functions $\psi_{X_i}$ for $X_i$, outlier thresholds
$\theta^{+}_{X_i}$ and $\theta^{-}_{X_i}$ for $\psi_{X_i}$, representations $U_i$ for $X_i$, find
all meaningful outlier relationships  of
$U_i, U_j, 1 \leq i \neq j \leq n$.
 
\noindent Note that the choice of outlier detection functions depends on a combination of domain and application needs, but in practice the user will likely use only 
a few functions over all attributes $X_i$. Our problem statement, however, does not restrict the number of outlier detection functions for flexibility, and also 
because outlier detection per se is outside the scope of this paper. 

\hide{ Given a collection of data sets
$\mathcal{D}$ and the union of their numerical attributes $\{X_i, 1 \leq i \leq n\}$, 
find all meaningful outlier alignments of $X_i, X_j, 1 \leq i \neq j \leq n$.}

\section{Discovering Meaningful Outlier Relationships}
\label{general-components}

\begin{figure}[t!]
\centering 
\includegraphics[width=0.8\columnwidth]{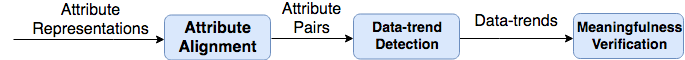} 
\caption{High-level overview of \tipo.}
\label{fig:diagram}
\end{figure}

In what follows, we introduce \tipo, a method designed to
identify meaningful relationships between outliers. 
%
Figure~\ref{fig:diagram} provides a high-level overview of \tipo.
The method takes as input a set of \emph{attribute representations}
derived by outlier detection functions 
(Definition~\ref{def:outlier-detection-function}) and associated
timestamps. Different functions can be used as long as they
appropriately rank the outliers and normalize values.
\emph{Attribute Alignment} derives a set of attribute pairs by joining the
different attributes on their timestamps.
Given a pair of attribute representations $U_1$ and $U_2$,
\emph{\Trend Detection} verifies whether the aligned scores of 
$U_1$ and $U_2$ form a \trend.
\hide{Here, we introduce \trends, a statistical model that captures the interactions
across aligned values, and are formally defined in
Section~\ref{subsec:regression}.}
If a \trend is detected, \emph{Meaningfulness Verification} checks 
whether it helps predict the relationship across aligned outliers. 
%
We describe each of these components in detail below. 

\subsection{Aligning Attributes}
\label{subsec:pruning}

There are a number of challenges involved in identifying outlier
alignments across different attributes. First, since
different attributes have values in different scales, metric
distortions may occur making it difficult to compare the values. While this can be addressed by using
normalization functions such as z-scores~\cite{CRR73} for the selected
outlier detection function $\psi_X$ (Definition~\ref{def:outlier-detection-function}),
these functions have a serious limitation for our problem scenario:
since they do not take cumulative effects over time into account,
they restrict the alignment of outliers to those that occurred at the
same timestamp.

\begin{figure}[t]
\centering
\includegraphics[width=0.8\columnwidth]{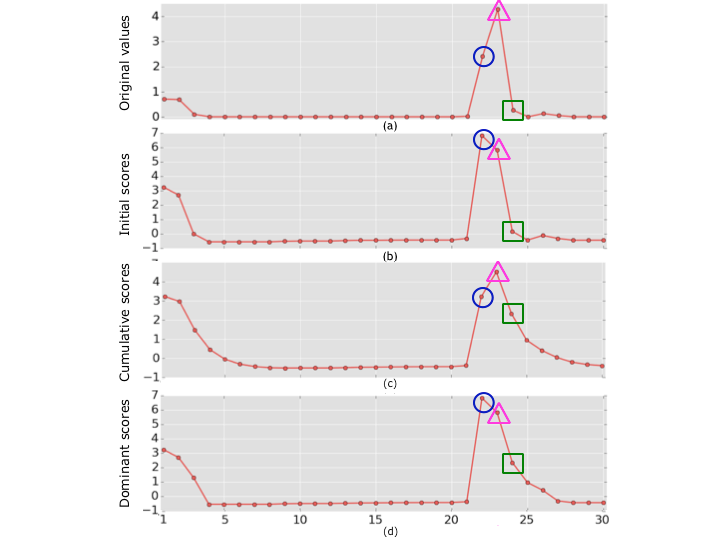} 
\caption{Different representations of daily average pluviometry levels
  in New York City in April, 2012. (a) Original values. (b) Variation
  of z-scores, where the mean and standard deviation related to each
  value are calculated over its previous 30 days.  (c) Cumulative scores
  ($\lambda = 0.5$) associated with the scores in (b). (d) Dominant scores
  associated with the scores in (b) and (c). Values enclosed
  by blue circles correspond to day 22; those enclosed by pink
  triangles, to day 23; and those enclosed by green rectangles, to day
  24.\hide{Would it be possible to have (b),(c) and (d) be on the same scale?}}
\label{fig:pluviometry_example}
\end{figure}

Consider Figure~\ref{fig:pluviometry_example}(a), which shows average
pluviometry levels in NYC for April, 2012.  Note that it rained
considerably on the $22^{nd}$ and on the $23^{rd}$ (blue circle and
pink triangle respectively), but not much on the $24^{th}$ (green
rectangle). This abrupt break in the pluviometry pattern is captured
by standard z-scores, 
as illustrated in Figure~\ref{fig:pluviometry_example}(b).
In practice, however, the heavy rain registered on the $22^{nd}$ and on the
$23^{rd}$ may help understand events in other data sets, even if they
happened a few days later due to the lingering effects of the heavy rain
such as flooding. Thus, to better understand interactions between temporal data sets, 
it is important to take the cumulative impact of events into account.

One alternative to address this problem  
would be to align
scores with different timestamps. This solution, however, is costly,
as it significantly increases the number of possible alignments. More
importantly, this strategy is not trivial to tune: how large should
the temporal range for score alignments be? And should this temporal
range depend on the initial score values?
In Section~\ref{related-work}, we discuss techniques that can be
used to address these issues and their associated performance
implications.
 
\paragraph{Capturing Cumulative Effects.}
We propose a new technique that augments scores to capture cumulative effects from outliers.
Intuitively, by boosting the score associated with the $24^{th}$
(Figure~\ref{fig:pluviometry_example}(c)), the chance of aligning it
with outliers from other data sets that capture the effects of the
heavy rain (e.g., data sets whose outliers happened a few days later)
would increase, thus naturally expanding the \textit{explanation
  power} of our approach.
The boosting strategy computes a \textit{cumulative score} for each
timestamp and compares it with the original score given by
$\psi_X$. The larger (absolute) score, referred to as \textit{dominant
  score}, is then used to represent the original value.
Alignments based on dominant scores are effective at capturing relevant cumulative effects, as we 
discuss in Section~\ref{subsec:effectiveness-baselines}. Furthermore, they are efficient to compute because two dominant scores 
have to share \textit{the same timestamp} in order to be aligned (see Definition~\ref{def:aligned-scores})---i.e., extra alignments between scores that are 
only temporally close,  which might be quadratic on the number of scores in the worst case, are never computed. 


Given a raw attribute $X$, the computation of dominant scores requires
just an initial outlier detection function $\psi_X$. 
Cumulative effects are then \textit{constructed iteratively} and, after a series of comparisons, the dominant scores
are generated. 
Consider, for instance, Figure~\ref{fig:pluviometry_example}(d), which
shows dominant scores as a possible representation for the values
in Figure~\ref{fig:pluviometry_example}(a).
These dominant scores are derived after comparisons with z-scores ($\psi_X$) in Figure~\ref{fig:pluviometry_example}(b).
Note that, for most points, the differences between the z-scores in
Figure~\ref{fig:pluviometry_example}(b) and the dominant scores in
Figure~\ref{fig:pluviometry_example}(d) are minor. However, if preceding
values are considerably large, as is the case for day 24 (green
rectangles), the dominant score significantly differs from the
z-score, carrying the notion of a cumulative effect.
%
%
Before introducing dominant scores, we need to formally define cumulative scores. 

\begin{define}[\textbf{Cumulative Score}] Let $\lambda \in [0,1]$ denote a
  fixed coefficient that regulates to which extent cumulative effects
  are taken into account, and $c^i \in \mathbb{R}$ be the
  \textit{cumulative score} associated with an initial score $u_X^i$
  determined by function $\psi_X$. Score $c^i$ is computed by the
  recursive function $\sigma: \mathbb{R} \rightarrow \mathbb{R}$:

  \begin{equation}
    c^i = \sigma(u_X^i| \lambda, c^{i - 1}) =
   \begin{cases}
     u_X^i, & \text{if $i = 0$}\\
     (1 - \lambda)u_X^i + \lambda c^{i - 1}, & \text{otherwise}\\
   \end{cases}
   \label{eq:cumulative_scores}
  \end{equation}
\end{define}

\noindent Cumulative scores are calculated in a tunable form, depending on a parameter $\lambda$ that can be set by the user.\footnote{Alternatively, a value for 
$\lambda$ could be learned, but this is outside the scope of this work.} \tds{We decided to model the lingering effect of scores in a tunable fashion because this effect could 
 vary depending on the context, as well as on the scale at which measurements are taken (e.g., seconds \textit{versus} days). In particular, in contexts where there are no lingering effects --- or where they are not relevant ---,
 $\lambda$ should be set to zero, leading to cumulative scores that are identical to the respective initial scores.} 
Note that cumulative scores are equivalent to exponential moving averages~\cite{roberts1959}, though in the latter $\lambda$ and $(1 - \lambda)$ are typically switched. 
We decided to model cumulative effects with exponential decay because it is a well understood weighting mechanism that fades quickly with time.  
 This property is important, as it increases the alignments only between outliers that are not too far apart.
It is worth noting that the fact that $\lambda \in [0, 1]$
guarantees the decay of cumulative effects. In other words, the impact
of a value in $t_i$ is carried over, but it decays to a
negligible amount within a few steps. 
Finally, note that, depending on the values of $\lambda$ and $c^{i - 1}$, $|c^i|$ may be larger than 
$|u_X^i|$. In these cases, we consider that \textit{the cumulative effect over $t_i$ is a better representation for $x^i$ than score $u_X^i$}.
%

Cumulative scores can, however, have an undesirable consequence:
they may dampen outliers. In Figure~\ref{fig:pluviometry_example}(c),
for example, the cumulative score for day 22 (blue circle) is
significantly smaller than its corresponding initial score (blue
circle in Figure~\ref{fig:pluviometry_example}(b)).  This happens
because previous pluviometry levels have very little cumulative impact
on the level of day 22.
As our goal is the discovery of meaningful relationships across
aligned outliers, we cannot use a scoring method that dampens 
them out. 
To address this limitation, we introduce the notion of dominant
scores, which take both initial and cumulative scores into account
without dampening either.

\begin{define}[\textbf{Dominant Score}] 
The \textit{dominant score}
  $u_{\delta}^i$ between $u_X^i$ and $c^i$ is computed by the function
  $\delta: \mathbb{R}^2 \rightarrow \mathbb{R}$:

  \begin{equation}
    u_{\delta}^i = \delta(u_X^i, c^i) = 
    \begin{cases}
      u_X^i, & \text{if $max(|u_X^i|, |c^i|) = |u_X^i|$}\\
      c^i, & \text{otherwise}\\
    \end{cases}
    \label{eq:dominant_score}
  \end{equation}
\end{define}

\noindent Given outlier thresholds $\theta^{+}_X$ and $\theta^{-}_X$, the number of
outliers generated with function $\delta$ (\textit{dominant outliers})
is \textit{never smaller} than the number of outliers generated with
function $\psi_X$.  After all, if $u_X^i$ is an outlier, we only have
$u_{\delta}^i = c^i$ if either $c^i > u_X^i > \theta^{+}_X$ or $c^i < u_X^i < \theta^{-}_X$. 
Using dominant scores thus guarantees that no values are dampened. For
example, note that every initial peak in
Figure~\ref{fig:pluviometry_example}(b) is represented as a peak in
Figure~\ref{fig:pluviometry_example}(d), which corresponds to the
dominant scores.

\paragraph{Alignment Index.}
After the dominant scores are generated, the next step is to align
attributes that have \textit{at least one pair of aligned dominant
  outliers}.  Note that, despite requiring a common timestamp, the
alignment between dominant outliers implicitly allows the matching of
abnormal events that do not occur at the same time, as their effects
get carried over.

%
Verifying which pairs of attributes have aligned dominant outliers can
be prohibitively expensive for data set collections that contain a
large number of attributes. Furthermore, most data set combinations
will have no aligned outliers.
To make the alignment process efficient and scalable, we introduce an
\textit{alignment index}, which can be created at the same time  the dominant
scores are computed, thus incurring no additional overhead. 

\begin{figure}[t!]
    \centering
    \includegraphics[width=0.6\columnwidth]{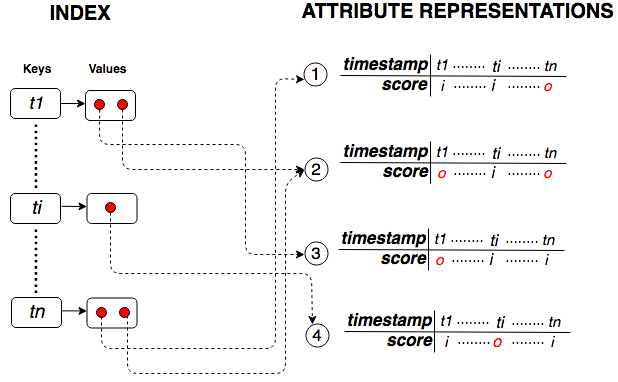}
    \caption{The keys of the index are timestamps $t_1, \ldots,
      t_n$.
      The attributes are represented by circled numbers 1,
      2, 3, and 4. Outliers are represented as $o$
      and inliers as $i$. The index values  connect
      an attribute to a timestamp if an outlier
      occurred in the timestamp in question.} 
\label{fig:index_structure}
\end{figure}

Each key in this index corresponds to a timestamp and is associated
with all attributes with dominant outliers for the timestamp in
question. 
The structure of the index is illustrated in
Figure~\ref{fig:index_structure}.

\begin{figure}[t!]
    \centering
    \includegraphics[width=0.5\columnwidth]{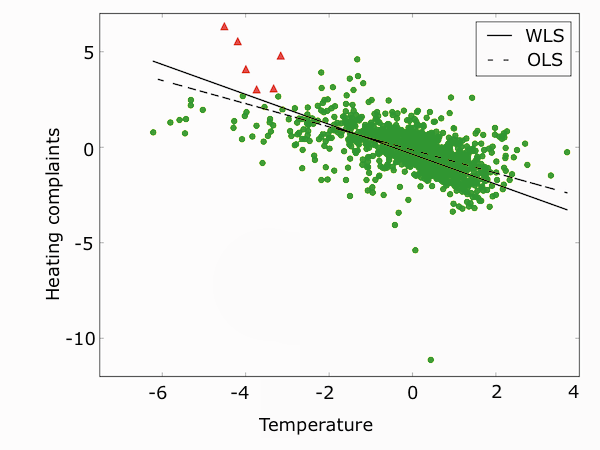}
    \caption{Dominant alignments between temperatures and heating complaints for 2011-2013.
      The function $\psi_X$ used in the construction of the dominant scores is a variation of z-scores,
      where the mean and standard deviation associated with a value are computed
      over the previous 30 days.  
      Red triangles represent aligned dominant outliers; green 
      circles, remaining alignments (outlier threshold $\theta_X = \theta^{+}_X = -\theta^{-}_X = 3$
      in both cases).  The regression models are built over all 
      alignments using $WLS$ and a weightless, ordinary linear regression ($OLS$).}
\label{fig:regions}
\end{figure}

Attributes that are associated with at least one index key in common,
i.e., that have at least one pair of aligned dominant outliers, are
retrieved for \textit{attribute alignment}.  
As we explain in Section~\ref{subsec:regression}, the alignment of all
possible (dominant) scores is critical for \trend detection.
For simplicity, we refer to alignments across dominant scores as  \textit{dominant alignments}.  
Each point in the scatterplot in Figure~\ref{fig:regions}, for example, corresponds to a dominant alignment between
heating complaints and temperatures.

Given the rarity of dominant outliers, and consequently of their
alignments, the use of the \emph{alignment index} substantially
reduces the number of pairs of attributes
compared. 
In practice, we expect that most 
attribute pairs have no aligned dominant outliers, thus being
naturally pruned away from the analysis. The impact of this index in
the performance of \tipo is discussed in
Section~\ref{subsec:scalability}.

\subsection{Detecting \Trends}
\label{subsec:regression}

%
Given the sets $U_1$ and $U_2$ containing the dominant scores for
attributes $X_1$ and $X_2$, respectively, our first task is to check whether there is a \trend across their aligned values. \Trends are defined as follows. 

\begin{define}[\textbf{\Trends}]
\label{def:corr-trend}
Given a set of pairs of aligned scores $\{(u_1^i, u_2^i)\}$, where $u_1^i \in U_1$, $u_2^i \in U_2$ and $i=1,...,N$,  
if there is a \textit{statistically significant linear regression model}~\cite{CRR73} across these pairs, we say that there
is a \textit{\trend}. 
\end{define}

Linear regression models have traditionally been used to detect linear
correlations, so they are a natural candidate for
the task of detecting interpretable trend patterns across aligned attributes. In
addition, these models have very low space complexity,
requiring the computation of only two parameters per regression line
(slope and intercept). Finally, they scale well with the number of aligned scores.
\tds{The use of kernels and other attribute transformations~\cite{hastie@statisticallearning} could be useful
in the detection of a larger pool of \trends, but it is outside the score of this paper.} 

To assess whether the computed regression line is statistically
significant, we verify if its slope is not statistically equivalent to
zero~\cite{CRR73}, as detailed later in this section.

\paragraph{Outlier-Biased Weighting Scheme.} Recall that our goal is to discover meaningful outlier relationships that
\textit{can be seen coming}. 
%
The intuition is that we expect non-random outlier alignments to follow a
pattern that is present in close-by aligned scores (near-outliers),
and that this pattern gets increasingly stronger for closer
near-outlier alignments.\footnote{Recall that close-by aligned scores are near in \textit{attribute space} --- not necessarily in time.} 
Consequently, given that \tipo captures such patterns with linear
regression models, it makes sense to calibrate them to prioritize the
fitting of near-outlier and outlier alignments.
%
%
For this purpose, we use \emph{weighted least squares}
(\textit{WLS})~\cite{CRR73} to estimate 
linear regression models. 

Without loss of generality, we assume that $U_1$ and $U_2$ have the
same size, and that any given score $u^i_{1} \in U_1$ is aligned with a
single score $u^i_{2} \in U_2$. 
We then model the relationship between aligned scores  
($u^i_{1}$, $u^i_{2}$) with a linear regression model $f_1$  of $U_2$ on $U_1$,
such that
\begin{equation}
\label{eq:regression}
\hat{u}^i_{2} = b_1u^i_{1} + a_1
\end{equation}
where $\hat{u}^i_{2}$ is the estimate of $f_1$ for $u^i_{2}$.
\textit{WLS}  allows 
distinct contributions, represented as weights for different pairs 
($u^i_{1}$, $u^i_{2}$), to the estimated model.  
It estimates $b_1$ and $a_1$ by minimizing the sum $S$ of
squared residuals, defined as
\begin{equation}
\label{eq:wls}
  S = \sum\limits_{i=1}^{|U_2|} w^i(u^i_{2} - \hat{u}^i_{2})^2 
\end{equation}
where weights $w^i$ are set a priori with a weighting scheme.

To model the 
 importance of scores based on their distance to outliers\hide{not clear...I would interpret this as distance to other outliers but I think the intent is to say extent of outlyingness or deviation? }, we propose a \emph{new outlier-biased weighting 
scheme for WLS}. 
\tds{Our goal is to allow partial weight to near-outliers as function of their deviance, while allowing equal weight to all outliers irrespective of their deviation, so that no single extreme outlier dominates the regression.} 
%
Let $u_X^i$ be the dominant score computed for attribute $X$ on timestamp $t_i$. 
The weight of $u_X^i$ is given by function
$\omega: \mathbb{R} \rightarrow \mathbb{R}$, defined as
\begin{equation}
\omega(u_X^i) =\begin{cases}
           1, & \text{if $u_X^i$ is an outlier}\\
           \alpha^{(\theta^{+}_X - u_X^i)}, & \text{if $0 \leq u_X^i < \theta^{+}_X$}\\
           \alpha^{(|\theta^{-}_X| - |u_X^i|)}, & \text{if $\theta^{-}_X < u_X^i < 0$}\\
           \end{cases}
\end{equation}  
\noindent where $\alpha \in (0, 1]$ is a  parameter fixed a priori.
\hide{ do we have examples of cases where neg. theta and pos. theta should be different? do we have an example where being generic here is useful? Why is it that,
for outliers, the weight is always 1? }\hide{see previous comment re why weight should be 1}

\begin{figure}[t]
    \centering
    \includegraphics[width=0.5\textwidth]{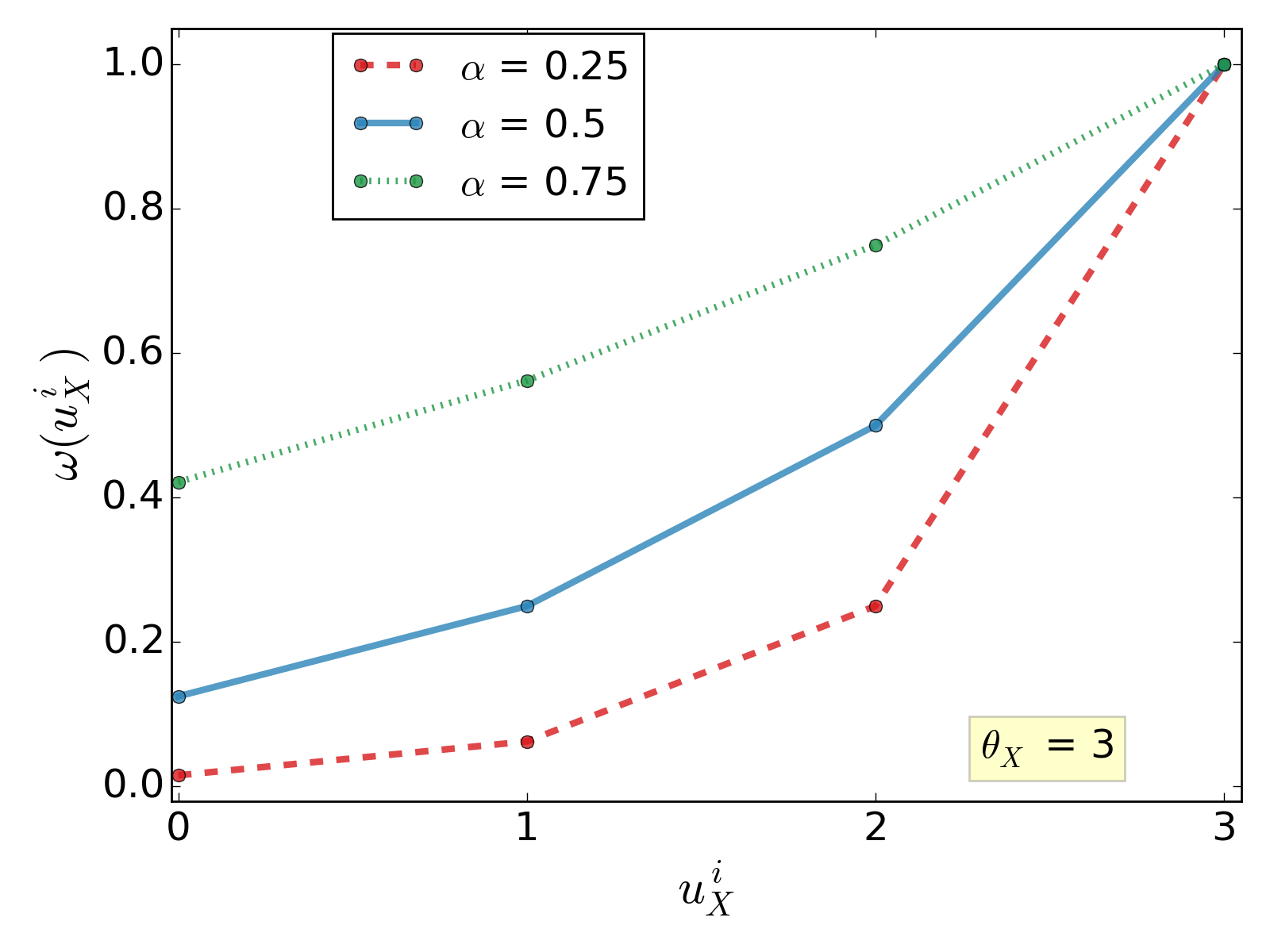}
\caption{Weight variations as parameter $\alpha$ changes. Scores $u_X^i$ are normally distributed around zero, and  $\theta_X = \theta^{+}_X = -\theta^{-}_X = 3$.}
\label{fig:varying-alpha}
\end{figure}

  
%
Consider Figure~\ref{fig:varying-alpha}, which shows how function
$\omega$ varies for different $\alpha$ values. In this example, we have a single outlier threshold $\theta_X = \theta^{+}_X = -\theta^{-}_X$ for simplicity. 
 Note that, as $u_X^i$ increases, weights $\omega(u_X^i)$ get larger for \textit{higher}   
 $\alpha$ values.  
Moreover,  
if $u_X^i$ is not an outlier, the value of function $\omega$ always gets larger as 
$|u_X^i|$ gets closer to $\theta_X$, irrespective of  $\alpha$.  
Finally, given $U_1$ and $U_2$, weight $w^i$ depends on aligned values ($u^i_{1}$, $u^i_{2}$), and is defined as 
\begin{equation}
  w^i = min(\omega(u^i_{1}), \omega(u^i_{2}))
\end{equation}

To illustrate the advantages of using \textit{WLS} with the proposed weighting scheme, 
consider the solid line in Figure~\ref{fig:regions}. 
By calibrating the contribution of scores differently, the distance
from the outliers to the \textit{WLS} line is smaller than the
distance to the dashed line, which was built with an equally-weighted\hide{\ab{equally-weighted? i.e., no differential weighting. }}, ordinary linear regression. 
The \textit{WLS} line  fits
the aligned outliers better by design.
Moreover, as the figure indicates, this  model adequately captures the pattern present in near-outliers.

\paragraph{Checking for \Trends.} 
The model represented by $f_1$ is determined by minimizing the sum $S$ of squared residuals for $U_2$ (Equation~\ref{eq:wls}). Note that 
 coefficients $b_1$ and $a_1$ would be different if this equation minimized residuals for $U_1$. The formulation above is thus asymmetric, posing the following 
limitations: (i) the order of $U_1$ and $U_2$ matters in the 
 investigation of their aligned outliers, which is counterintuitive because relationships between outliers do not have a preferred direction, i.e., outliers in $U_1$ 
 may help understand outliers in $U_2$ \textit{and} vice-versa; and (ii) an asymmetric formulation can confuse end-users, leading to the belief that outliers of one 
 attribute are \textit{causing} outliers of another; here, we aim
 to find explanations through general associations and not causal relations. 
%
To address these limitations, we also create a weighted regression
model $f_2$ of $U_1$ on $U_2$, defined as
 \begin{equation}
\label{eq:symm-regression}
\hat{u}^i_{1} = b_2u^i_{2} + a_2 
\end{equation}
%
\noindent where $\hat{u}^i_{1}$ is the estimate of $f_2$ for $u^i_{1}$. Coefficients $b_2$ and $a_2$ are estimated with a minimization analogous to 
Equation~\ref{eq:wls}.

Given the regression models $f_1$ and $f_2$, we verify if there is a statistically significant linear relationship for \textit{at least one of them}. 
If there is no significant relationship, scores
in $U_2$ do not tend to change as scores in $U_1$ change, and
vice-versa.
This happens when coefficients $b_1$ and $b_2$ are not
statistically different from zero.  Thus, to detect whether there is a
 significant \trend between $U_1$ and $U_2$, we have to apply the
\textit{regression slope test}~\cite{CRR73}.  Given a slope $b$, the
test examines the following null ($H_0$) and alternative ($H_a$)
hypotheses: 

\begin{description}
\item $H_0$: $b = 0$
\item $H_a$: $b \neq 0$
\end{description}

The sampling distribution of slope $b$ is used to determine whether it
is statistically different from zero~\cite{CRR73}.  \hide{If \ab{DROP - we do the test
for slopes $b_1$ and $b_2$} and the null hypothesis gets rejected for
\textit{at least} one of them\ab{b1 or b2}}
If the null hypothesis gets rejected for \textit{at least} one of $b_1$ or $b_2$, component \textit{\Trend Detection} concludes
that there is a \trend across the aligned scores of $U_1$ and
$U_2$. 
As an example, consider the \textit{WLS} model of heating complaints
on temperatures in Figure~\ref{fig:regions}: the null hypothesis is
rejected due to the non-zero slope, i.e., this model corresponds to a
\trend.

\subsection{Verifying Meaningfulness}
\label{subsec:detection}

If a \trend is detected, associated with either 
regression model $f_1$ or $f_2$, component 
\emph{Meaningfulness Verification} 
checks if it models the data accurately and if it provides statistical evidence that the outlier relationship is  meaningful.
Without loss of generality, we assume throughout this section
that there is a \trend associated with regression model $f_1$ of $U_2$
on $U_1$. 
Initially, \emph{Meaningfulness Verification} checks whether $f_1$ has a \textit{reasonable goodness-of-fit},
defined as follows.

\begin{define}[\textbf{Reasonable Goodness-of-fit}] \label{def:reasonable-goodness-of-fit} 
Given a set of pairs of aligned scores of $U_1$ and  $U_2$, 
a linear regression model $f_1$ of $U_2$ on $U_1$ has reasonable goodness-of-fit~\cite{CRR73} if \hide{its coefficient of determination $\bar{R}^2$ is above a given threshold $\bar{R}^2_{min}$. if} a fraction of at least $\bar{R}^2_{min}$ of
the variance of the aligned scores is explained by it. \hide{--- could you rephrase please? if the R2 is above some threshold?} 
\end{define} 
\noindent A reasonable goodness-of-fit is important because \trends that are statistically significant may still capture random patterns that are not 
faithful to the data. 
In other words, if the discrepancy between values
$\hat{u}^i_{2}$ and $u^i_2$ is too high, $f_1$ is not reliable enough
to help assess the meaningfulness of outlier relationships.  
In practice, we assess the goodness-of-fit of a \trend by computing its
\textit{adjusted r-squared}, $\bar{R}^2$~\cite{CRR73}. 
%
\tds{As an example, consider the \textit{WLS} model in Figure~\ref{fig:regions}, whose $\bar{R}^2$ is approximately 0.67.  
If $\bar{R}^2_{min} = 0.50$, for 
instance, this model would have a reasonable goodness-of-fit. }




If the \trend associated with $f_1$ has a reasonable goodness-of-fit, \tipo verifies if the aligned outliers are \textit{consistent} with it.
Formally, let $O_1 \subset U_1$ and $O_2 \subset U_2$ be subsets comprising all outliers of $U_1$ and $U_2$ respectively.  
 For simplicity, assume that $O_1$ and $O_2$ have the same size and any given score $o^i_{1} \in O_1$ is aligned with a score $o^i_{2} \in O_2$.
 Every point in Figure~\ref{fig:regions}, for example, corresponds to an alignment between scores of $U_1$ and $U_2$. The red triangles in particular 
 correspond to the aligned scores of $O_1$ and $O_2$.  
 With model $f_1$, one can generate estimates $\hat{o}^i_2$ for scores $o^i_{2}$.  
 To verify the consistency of the aligned outliers with respect to $f_1$, let $E_O$ be the error distribution of estimates $\hat{o}^i_2$ with 
respect to scores $o^i_2$,  
$E_U$ be the error distribution of estimates $\hat{u}^i_2$ with respect to scores $u^i_2$ (used to build model $f_1$), 
 $\rho$ be a high percentile of $E_U$, such as the $95^{th}$ percentile, and $0.5 \leq \beta \leq 1$ be a threshold percentage.  

\begin{define}[\textbf{Consistency}] \label{def:consistency} 
The aligned outliers of $O_1 \subset U_1$ and $O_2 \subset U_2$ are \textit{consistent} with a \trend associated with $f_1$ 
if \textit{at least} a fraction $\beta$ of the errors in $E_O$ are bounded by $E_U$'s percentile $\rho$.  
\end{define}

\noindent Since trend behaviors are usually associated to a certain level of
noise, our definition of consistency needs to be resilient and 
allow for a few exceptions, i.e, a few aligned 
outliers that may not be in keeping with the \trend behavior. 
This is the intuition behind $\beta$, a threshold that ensures that the majority of the aligned outliers has to be consistent with $f_1$. 
As for $\rho$, it has to be a high percentile of $E_U$ in order to guarantee a high level of significance. 
Finally, \tipo uses the Euclidian distance to calculate error distributions $E_O$ and $E_U$ because it is a common, easy to interpret
choice~\cite{CRR73}. It is, however, possible to apply different error functions.  

\tds{Consider again Figure~\ref{fig:regions}. If we set $\rho$ as the
$95^{th}$ percentile of $E_U$, we have that approximately 83\% of the
errors for the outlying heating complaints generated with the
\textit{WLS} model are bounded by $\rho$.  
Consequently, if $\beta = 0.67$, the aligned outliers are 
consistent with this model.}
In practice, distribution $E_U$ may be too small for reliable $\rho$
percentiles.  We thus use the bootstrap method (based on sampling with replacement) for estimating the
percentile and its confidence 
interval~\cite{EfroTibs93}.

If $f_1$ has a reasonable goodness-of-fit and the aligned outliers are
consistent with it, $f_1$ is evidence that the relationship across the aligned outliers is 
meaningful. Depending on how parameters are set, this is the case in 
the \textit{WLS} model in Figure~\ref{fig:regions}, for example.
Component \textit{Meaningfulness Verification} verifies goodness-of-fit and consistency for
\textit{every model with a \trend}, i.e., if $f_1$ and
$f_2$ are associated to \trends, both are evaluated. Aligned 
outliers just need to be consistent with one reasonable model (either 
$f_1$ or $f_2$) to be considered statistically significant.

\subsection{Putting it all Together}
\label{subsec:putting-together}

Algorithm~\ref{alg:general_structure} details how the different steps 
described above are combined to discover meaningful outlier relationships. 
%
\tds{Given a collection of data set attributes, the goal is to output every pair for which there is a meaningful outlier relationship. To this end, we first represent attributes with dominant scores and  prevent unnecessary attribute
comparisons by using an index. 
Next, a search for \trends is performed for every attribute pair that has at least one outlier alignment. Whenever a \trend is detected for a pair, the search is followed by an analysis of
 statistical meaningfulness and, when the underlying outlier relationship is considered meaningful, the attribute pair becomes a part of the algorithm's output.}  
%

%
%
%
%

\RestyleAlgo{ruled}
\SetAlFnt{\small}

\begin{algorithm} 
\SetAlgoLined
\KwIn{\textbf{\tipo parameters:} Coefficient $\alpha$, thresholds $\beta$ and $\bar{R}^2_{min}$}
\KwIn{\textbf{Application parameters:} Attribute representations $U^{Init}_1$, \ldots, $U^{Init}_N$, coefficient $\lambda$,
  outlier thresholds $\theta^{+}_1$, $\theta^{-}_1$, \ldots, $\theta^{+}_N$, $\theta^{-}_N$}
\KwOut{Set \textit{detected}}

\textit{$Indexed\_Pairs$} $\gets \textit{AttributeAlignment($U^{Init}_1$, \ldots, $U^{Init}_{N}$, $\lambda$, $\theta^{+}_1$, $\theta^{-}_1$, \ldots, $\theta^{+}_N$, $\theta^{-}_N$)} $\;
\textit{detected} $\gets \varnothing$\;
\For{\textit{each pair ($U_i$, $U_j$) in $Indexed\_Pairs$}}{ 
   \textit{$f_1$, $b_1$\_passed, $f_2$, $b_2$\_passed} $\gets \textit{\Trend
Detection ($U_i$, $U_j$, $\alpha$, $\theta^{+}_1$, $\theta^{-}_1$, \ldots, $\theta^{+}_N$, $\theta^{-}_N$)} $\;
    \textit{meaningful} $\gets False$\;
    \If{\textit{$b_1$\_passed = True}}{
      \textit{meaningful} $\gets \textit{MeaningfulnessVerification($f_1$, $U_i$, $U_j$, $\beta$, $\bar{R}^2_{min}$, $\theta^{+}_i$, $\theta^{-}_i$, $\theta^{+}_j$, $\theta^{-}_j$)} $\;
    }
    \If{\textit{meaningful = False} and \textit{$b_2$\_passed = True}}{
      \textit{meaningful} $\gets \textit{MeaningfulnessVerification($f_2$, $U_j$, $U_i$, $\beta$, $\bar{R}^2_{min}$, $\theta^{+}_i$, $\theta^{-}_i$, $\theta^{+}_j$, $\theta^{-}_j$)} $\;
    }
    \If{\textit{meaningful = True}}{
      \textit{detected} $\gets$ \textit{detected} $\cup \{(U_i, U_j)\}$\;
    } 
 }

\KwRet{\textit{detected}}
\caption{General Structure of \tipo}\label{alg:general_structure}
\end{algorithm}

\paragraph{Algorithm Input - \tipo Parameters} The algorithm receives values for $\alpha$, $\beta$, and $\bar{R}^2_{min}$ as inputs, and they tune 
the behavior of \tipo concerning \trend detection and verification of statistical meaningfulness.   
As discussed in Section~\ref{subsec:regression}, the closer parameter $\alpha$  is to 1, 
the more similar the weights of distinct points are.~\footnote{Note that if $\alpha = 1$, the linear regression becomes ordinary, i.e., the weight of every point 
is the same.} 
The value for $\beta$ depends on how rigorous the requirements of the
application are: if $\beta = 0.67$, for instance, at least 67\% of the
aligned outlier errors have to be bound by the model's error
distribution.
As for $\bar{R}^2_{min}$, its value determines the minimum adjusted r-squared above which \tipo considers a model 
reasonable. Note that suitable values for $\bar{R}^2_{min}$ depend on the context: if the relationship between two aligned 
attributes is expected to be strong, $\bar{R}^2_{min}$ can be higher; if weak to moderate relationships are expected to be the majority, 
which is common for example in social science scenarios, $\bar{R}^2_{min}$ can be lower~\cite{cohen@psychbulletin1992}.
%
%

\paragraph{Algorithm Input - Application Parameters} Other input parameters include: initial attribute representations 
computed with compatible representation functions $\psi$, coefficient
$\lambda$ for the computation of dominant scores, and suitable outlier
thresholds. 
\tds{These parameters do not modify the behavior of \tipo (i.e., they do not tune any of its components), and  the user sets them according to the application. 
%
For example, if the user is interested in relationships across strictly co-occurring outliers, cumulative effects do not have to be modeled 
and $\lambda$ should be set to zero (see Equation~\ref{eq:cumulative_scores}). Moreover, if attributes can be well approximated by Gaussian distributions, the user can  
compute z-scores as functions $\psi$ and use outlier thresholds $\theta^{+}_X = -\theta^{-}_X = 3$. Alternatively, if the analysis only concerns positive outliers 
(e.g., very long wait times in doctor's offices) and values are normally distributed, the user can work with thresholds $\theta^{-}_X = -\infty$ and $\theta^{+}_X = 3$.} 
Note that the selection of these parameters, which are core components in the context of outlier
detection, are \textit{independent of \tipo}.  After all, the goal of
our method is the detection of \textit{meaningful outlier relationships}, not
the detection of outliers. 

\paragraph{Structure of Algorithm} Algorithm~\ref{alg:general_structure} first invokes \emph{Attribute Alignment} to produce
attribute pairs ($Indexed\_Pairs$) that have at least one dominant outlier alignment
(\lineref{1}) and whose values represent dominant scores.
\tipo then searches these pairs in which the \trend across aligned outliers is statistically significant (\emph{lines 3--15}). 
The first step in the loop executes \emph{\Trend Detection}~(\lineref{4}) to compute weighted regression models $f_1$
and $f_2$ over $U_i$ and $U_j$.
If the slope $b_1$ of model $f_1$ is statistically different from
zero, the test in \lineref{6} succeeds.
The component \emph{Meaningfulness Verification} then verifies whether
the aligned outliers of $U_i$ and $U_j$ are consistent with model
$f_1$ (\lineref{7}). If so, pair $(U_i, U_j)$ is added to
\textit{detected}, i.e., the co-occurring outliers of $U_i$ and $U_j$
classified as meaningful (\emph{lines 12--14}).
Otherwise, if the slope $b_2$ of model $f_2$ is statistically
different from zero, component \textit{Meaningfulness Verification}
is executed again to perform a similar error analysis for model $f_2$
(\emph{lines 9--11}). 
This second check is important because 
\trends
associated to $f_1$ and $f_2$ are equally relevant for the detection
of meaningful outlier relationships.
Note that outlier thresholds are passed to the three major components --- 
technically, all of them require a distinction between inliers and
outliers.  In practice, however, this process is optimized and the
separation between inliers and outliers is performed only once.

\section{Experimental Evaluation}
\label{setup-and-experiments}
We performed an experimental evaluation to assess the effectiveness of \tipo at detecting 
meaningful outlier relationships. 
We quantitatively assess different components of our solution, notably:
the usefulness of dominant scores, the benefits of taking cumulative
effects into account, the effectiveness of the \textit{Outlier-Biased} weighting scheme, and the efficiency
gains attained by the \emph{alignment index} as the number of 
attributes considered increases.
In addition to that, we evaluated \tipo qualitatively with case studies. 
%
Finally, we carried out a sensitivity analysis for parameters
$\alpha$, $\beta$, $\bar{R}^2_{min}$, and $\lambda$. 
%

One challenge we faced in the evaluation was that since \tipo is the
first approach for discovering meaningful outlier relationships, there
were no existing benchmarks we could use.  Therefore, as
described in this section, we had to create gold data with input from
human annotators.

\subsection{Experimental Setup}
\label{subsec:experimental-setup}

\paragraph{Data Sets and Attributes.} We used data sets provided by
different New York City agencies and the National Oceanic and
Atmospheric Administration (NOAA).
%
Table~\ref{tab:data-sets} lists the data sets, some of their properties, and examples of 
attributes used in our experiments. 
We aggregated the records in these data sets over days, and 
before aggregation their sizes varied from MBs to GBs.\footnote{The data sets are available at
  \url{https://figshare.com/collections/New_York_City_s_Urban_Data/4273586}. 
  The \textit{Taxi} data can be obtained from
  \url{https://data.cityofnewyork.us}.}

\begin{table}[t!]
\centering
\small
\begin{tabular}{p{1.2cm}p{3.5cm}p{1.2cm}p{1.2cm}p{1.2cm}p{3.5cm}}
\toprule
\textbf{Data Set} & \textbf{Description} & \textbf{Size} & \textbf{Year Range} & \textbf{Agency} & \textbf{Attribute Examples} \\
\midrule
\textit{311} & Requests to NYC's non-emergency complaint service 311 & 1.3GB & 2010-2018 & 311 & Number of complaints in categories: noise, heating, street condition\\ 
\midrule
\textit{Citi Bike} & Data from NYC's bike sharing system & 7.9GB & 2013-2017 & CitiBike & Number of trips\\ 
\midrule
\textit{Collisions} & Vehicle collision data & 254MB & 2012-2018 & NYPD & Numbers of: collisions, persons injured, cyclists killed\\
\midrule
\textit{Weather} & Meteorological data for NYC & 413MB & 2010-2018 & NOAA & Average: pluviometry level, temperature, wind speed\\
\midrule
\textit{Taxi} & NYC taxi trips & 170GB & 2009-2016 & TLC & Number of trips, average trip speed\\
\midrule
\textit{Crimes} & Criminal offenses reported to the New York City Police Department & 1.4GB & 2006-2015 & NYPD & Number of offenses\\
\midrule
\textit{Turnstile} & Turnstile data collected from NYC's subway system & 7.4GB & 2010-2018 & MTA & Number of subway trips\\
\bottomrule
\end{tabular}
\caption{Data set properties and attributes. In the experiments we
  used a total of 84 attributes. \hide{We did not use all numerical attributes available because some of them were noisy, redundant, or sparse.} The reported sizes
  refer to the original data sets before aggregation.}
\label{tab:data-sets}
\end{table}

\paragraph{Representation Function.} To derive the attribute
representations, we used a function based on \textit{mean residuals},
a simple variation of z-scores, as discussed below. We refer to this function as
$\psi_{MR}$.
The mean residual of a value captures how much it deviates from the overall expected behavior based on its \textit{past values}. 
 The function
is easy to implement, fast to compute, and suitable for both
streaming and static temporal
data~\cite{dasu@pvldb2015}. 
\begin{define}[\textbf{Mean Residuals}] 
Let $x^i \in X$, and $\phi \in \mathbb{N}^+$ be
the size of a time window $\kappa$ immediately preceding $x^i$,
corresponding to the interval $[t_i-\phi,t_i-1]$.
Moreover, let $\mu_{t_i, \phi}$ and $\sigma_{t_i, \phi}$ be
the mean and standard deviation of the values in $\kappa$. 
The mean residual $u^i$  of $x^i$ with respect to $\kappa$ is given by
$\psi_{MR}: \mathbb{R} \rightarrow \mathbb{R}$:
\begin{equation}
u^i = \psi_{MR}(x^i| \mu_{t_i, \phi}, \sigma_{t_i, \phi}) = \frac{x^i - \mu_{t_i, \phi}}{\sigma_{t_i, \phi}}
\end{equation}
\end{define}

\noindent Note that  $\psi_{MR}$ respects the requirements for representation functions presented in
Section~\ref{sec:problem-definition}. An advantage of using
$\psi_{MR}$ instead of standard z-scores is that different time window sizes capture different
kinds of outliers, from local to global: 
a small $\phi$
captures local patterns of the original values, 
while a large
$\phi$ captures global peaks and valleys in the raw values.
To capture a wider range of temporal effects, from local and
seasonal to global events, 
we experimented with 
time window sizes of 28, 30, 64, 90, 128, 180, 256, 360, and 365
days, totaling 756 different representations for the 84
attributes we considered.
%
It is important to point out that, despite the flexibility to capture
different types of outliers, mean residuals \textit{do not explicitly take
  cumulative effects from previous timestamps into account}. 

\paragraph{\tipo Parameters.} For the weighting scheme, we used a default value of $\alpha = 0.5$.  Since
$\alpha \in (0;1]$, choosing a value in the middle of the domain
generates weights that are neither too similar nor too different for
distinct points. 
We set $\rho$ as the $95^{th}$ percentile of the model error distribution  
and $\beta = 0.67$, i.e., we expect that at least $67\%$ of the
 errors for aligned outliers are drawn from the model error
distribution with a $95\%$ level of significance (see Definition~\ref{def:consistency}). 
The rationale behind this value of $\beta$ is the common modeling assumption that, under a fixed normalization, inliers come from the 
same distribution at a given point in time (an expected distribution), but outliers could come from a number of 
distinct, unexpected distributions. 
As a consequence, if most aligned outliers 
are consistent
with the computed \trend, then we have a strong indication that
\emph{the outliers can be seen coming}.
%
Finally, we  use $\bar{R}^2_{min} = 0.25$, considered a moderate value,
because in urban data or social science contexts, where there is a
plethora of intervening variables at play, such values are considered
appropriate~\cite{cohen@psychbulletin1992}.

\paragraph{Application Parameters.} To allow for a balanced contribution from both mean residuals and
cumulative scores, we set $\lambda = 0.5$. \hide{We discuss the
effectiveness achieved with other $\lambda$ values in
Appendix~\ref{app:lambda-variations}.}
For all experiments and gold data (described below), we use a single $\theta_X = \theta^{+}_X = -\theta^{-}_X = 3$ because 
this value is commonly used to isolate outliers  
in distributions that,
as is the case with both mean residuals and dominant scores, are
symmetric around the
mean~\cite{Aggarwal2013,Iglewicz1993,pukelsheim1994stats}. In such distributions, $\theta_X = 3$  corresponds to a high level of confidence (e.g., more than
99\% for Gaussian distributions) in the outliers. 

\paragraph{Standard Values.} For the regression slope test, we reject the null hypotheses at the standard 
significance level of 0.05.  
%
%

\subsection{Generation of Gold Data}
\label{subsec:gold-data}
%

To evaluate the effectiveness of
\tipo, we need to ascertain whether it is able to discover
meaningful outlier-based relationships. Since there are no benchmarks that
contain such relationships, we had to create gold data. 
To do so, we gathered attribute pairs with and without meaningful trend patterns across their aligned outliers. The meaningfulness of a trend pattern
was empirically assessed by human annotators, as explained below.

\begin{figure}[t!]
    \centering
    \subfloat[\textit{clear positive}]{\includegraphics[width=0.35\textwidth]{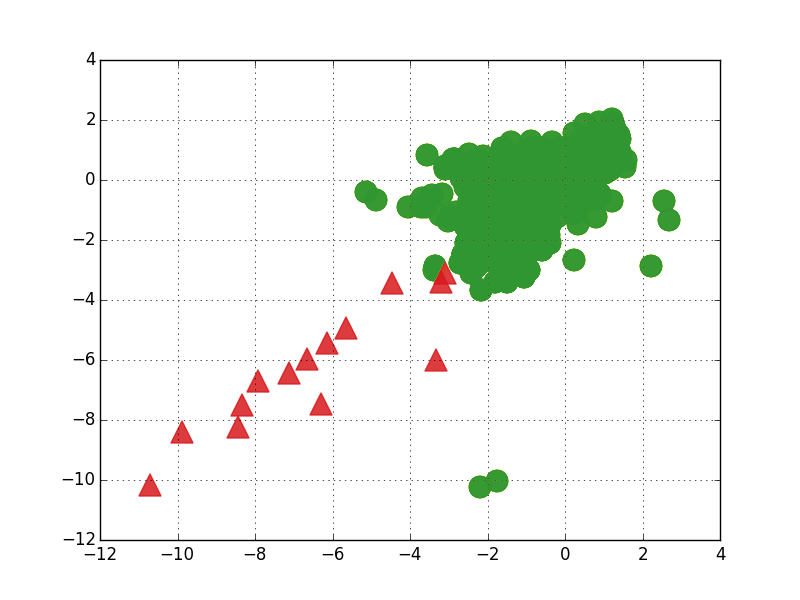} \label{subfig:easy_positive}}
    \qquad
    \subfloat[\textit{dubious positive}]{\includegraphics[width=0.35\textwidth]{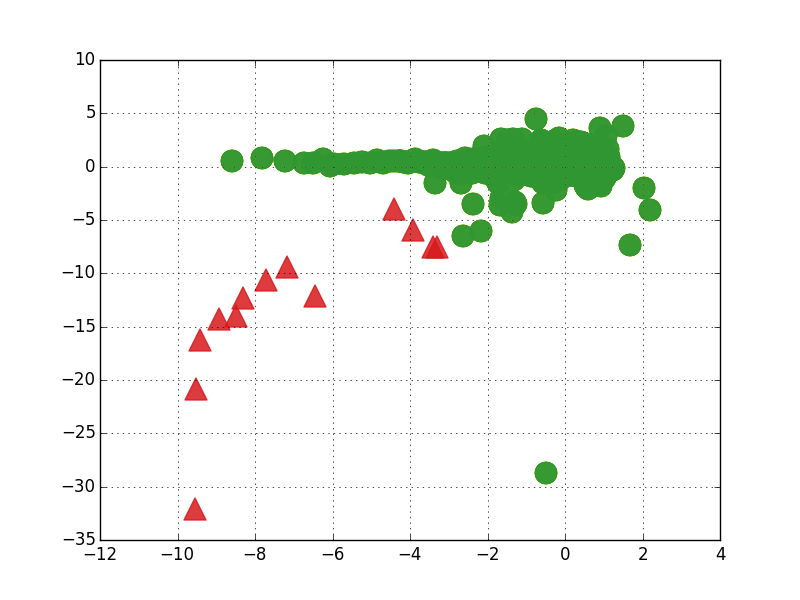} \label{subfig:hard_positive}}
    \qquad
    \subfloat[\textit{clear negative}]{\includegraphics[width=0.35\textwidth]{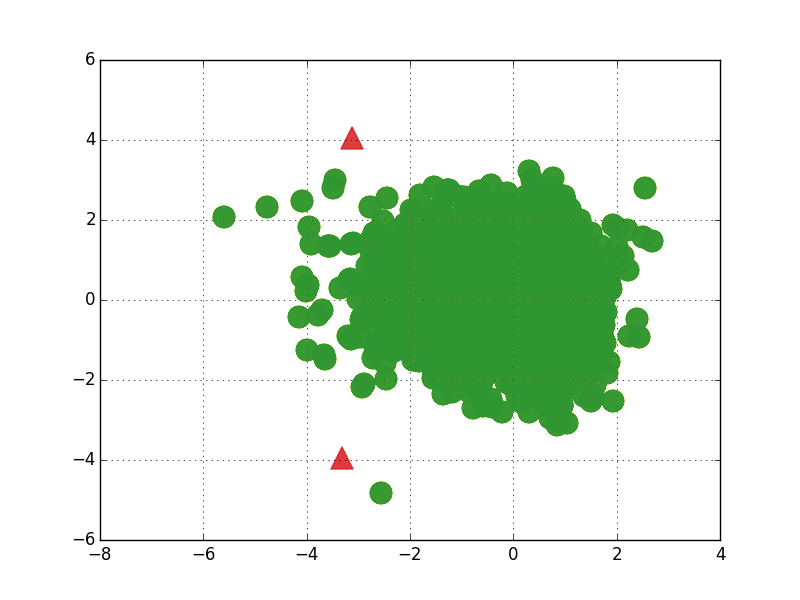} \label{subfig:easy_negative}}
    \qquad
    \subfloat[\textit{dubious negative}]{\includegraphics[width=0.35\textwidth]{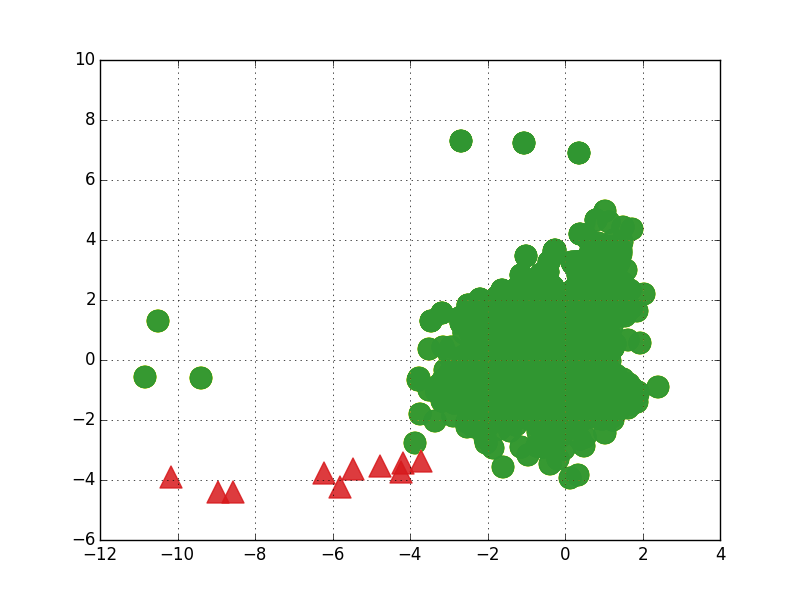} \label{subfig:hard_negative}}
\caption{Examples for the different labels in the gold data.}
\label{fig:gold-data-examples}
\end{figure}

\paragraph{Annotating Examples.} After deriving mean residual representations for the attributes in the data sets listed
 in Table~\ref{tab:data-sets}, we derived their corresponding dominant scores.  Recall that dominant scores allow potential matches of
close-by outliers. 
The next step was the generation of scatterplots with aligned dominant scores for several randomly selected pair of attributes. 
The graphs in Figure~\ref{fig:gold-data-examples} are examples of the 
scatterplots we generated.
We selected 150 scatterplots and showed them to five annotators,
who examined the plots independently.\footnote{During the annotation
  process, we omitted the labels of the scatterplots' axes to avoid  bias in the annotation. In other words, by hiding the labels we guarantee that all 
annotators have the \textit{exact same} information about the data.}  
For each scatterplot, we asked the annotators to answer the following question: \textit{Do you
  see a meaningful pattern across the aligned outliers?} 
While annotators interpreted the term \textit{meaningful} in slightly
different ways, they intuitively focused on a few common aspects: the
presence of a reasonable number of aligned outliers; the 
existence of a trend pattern among the outliers; and a certain shape
coherence across aligned outliers and/or near-outliers. 
For each scatterplot, the corresponding attribute 
pair was thus annotated with one of the following labels:

\begin{itemize}
\item \textit{clear positive}, if \textit{all} annotators saw a
  meaningful pattern;
\item \textit{dubious positive}, if \textit{most} annotators saw a
  meaningful pattern, but some annotators disagreed;
\item \textit{clear negative}, if \textit{no} annotators saw a
  meaningful pattern;
\item \textit{dubious negative}, if \textit{most} annotators did not
  see a meaningful pattern, but some disagreed.
\end{itemize}

Figure~\ref{fig:gold-data-examples} provides examples for each label. Finally, we selected 100 pairs to obtain a balance across the
different types of alignments in the gold data: \textit{positive}
(25 \textit{clear} and 25 \textit{dubious} pairs), and
\textit{negative} (25 \textit{clear} and 25 \textit{dubious} pairs).
We ran \tipo over the gold data and evaluated its performance using
\textit{recall}, \textit{precision}, and
\textit{F-measure}~\cite{baeza2011IR}.

\begin{figure}[t!]
    \centering
    \includegraphics[width=0.6\columnwidth]{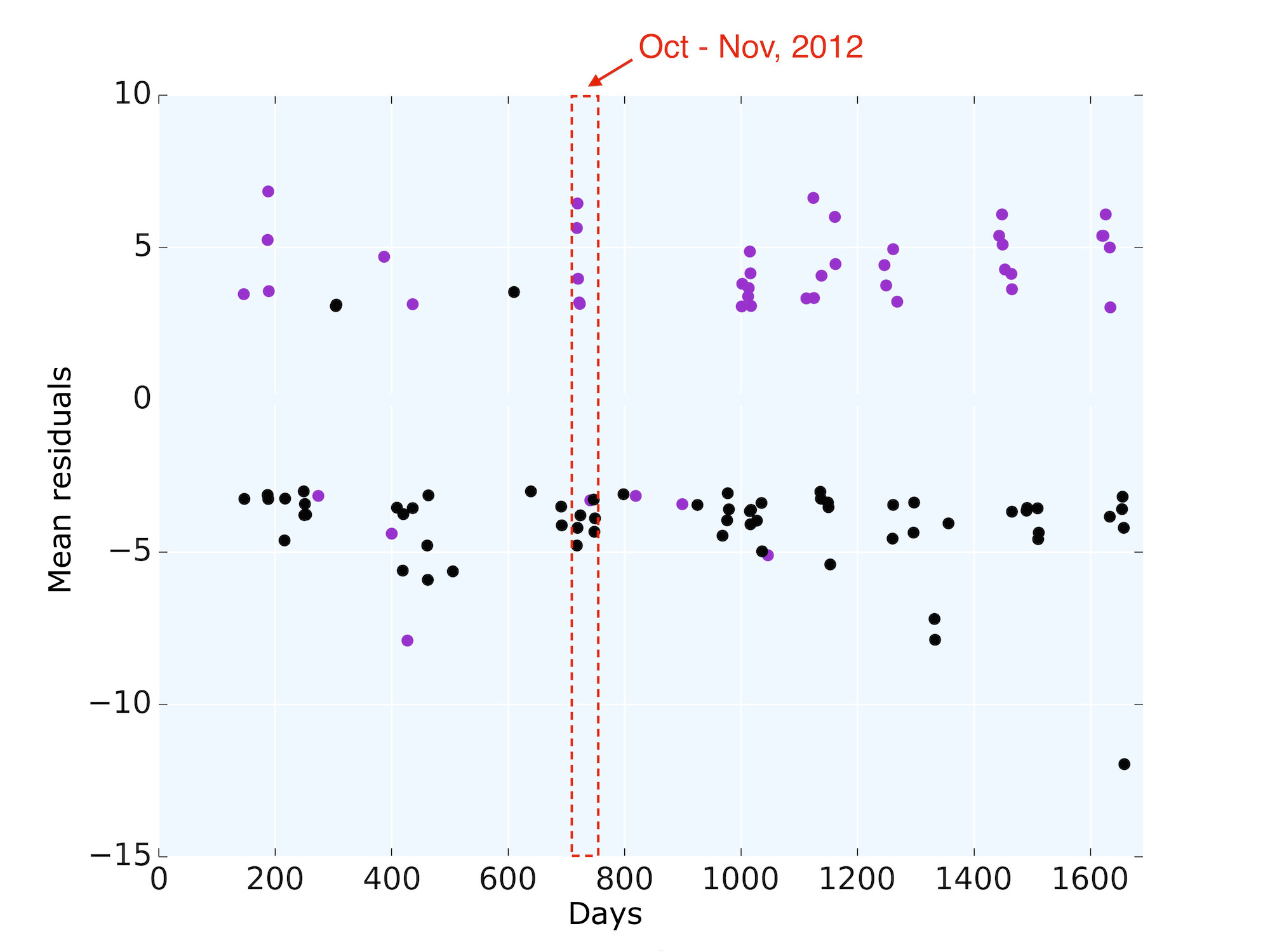}
\caption{Mean residual outliers for heating complaints (purple) and temperature (black) for years 2010-2017. In this example, we used window size $\phi = 30$ 
in the representation of  both attributes.}
\label{fig:outlier_time_series}
\end{figure}

\paragraph{Do outliers occur in close temporal proximity?} Before constructing the gold data we just described, we needed to check whether
 taking cumulative effects into account when aligning scores made sense in practice. We then checked, before generating dominant scores for 
several random pairs of mean residual representations, whether their outliers tended to occur approximately on the same
 days, i.e., if there is evidence that the incorporation of cumulative effects may help detect pairs of outliers that are slightly mismatched in time. 
We plotted mean residual outliers associated to different attributes, along with the days on which they occur, and verified whether such days are temporally close. 
As an example, consider Figure~\ref{fig:outlier_time_series}, which shows mean residual outliers
for heating complaints (purple) and temperature (black) over the course of several
years.  Note that these outliers generally occur in close temporal
proximity, often presenting a temporal mismatch of just a few
days.
After analyzing several such plots, 
we observed that slightly mismatched outliers are common.
This highlights the importance of modeling cumulative effects --- without it, it is not possible to align outliers that
 do not co-occur but may be semantically related. 

%

%

\subsection{Effectiveness: Quantitative Evaluation}
\label{subsec:effectiveness-baselines}

\paragraph{Baselines.}
To evaluate the different components of our approach and assess our
design decisions, we compared \tipo\footnote{We instantiated \tipo
  with the parameter values described in
  Section~\ref{subsec:experimental-setup}. In this section, every
  mention of \tipo and its results implicitly refers to this
  particular instantiation.} against a series of baselines. The first
three are constructed by varying components or fixing the parameters of
\tipo: 
\begin{itemize}
\item \textit{\tipo-Sub} - restricts models to aligned inliers, i.e.,
  values in interval $[-3; 3]$, enables us to assess the improvements obtained
from building models over \textit{all} aligned dominant scores;
\item \textit{\tipo-MR} - represents attributes with mean residuals, forgoing
  the use of cumulative effects, i.e., component \textit{Attribute Alignment} does not generate
  dominant scores and aligns mean residuals instead. This helps us understand the impact of taking
\emph{cumulative effects} into account; 
\item \textit{\tipo-OLS} - performs an ordinary linear regression by
  setting $\alpha = 1$, i.e., it does not weight the regression in
  order to better fit aligned outliers. By doing so, we can assess the
effectiveness of the weighting scheme in \textit{\Trend Detection}.  
\end{itemize}


\tds{Although traditional correlation metrics, such as Pearson and Spearman's rank correlations, are not adequate for small sample sizes 
because of their increased variability, and because their low statistical power might inflate the false positive rate~\cite{sample-size,power-failure}, 
we still decided to compare \tipo against them. The goal is to understand their limitations and potential qualities in practice, over concrete 
 samples of aligned outliers. To this end, we devised the following baselines:
\begin{itemize}
\item \textit{Pearson} - computation of the Pearson correlation coefficient over all aligned outliers; if the  coefficient is larger than 0.30 in modulus, 
and its p-value is lower than 0.05, the outlier relationship is considered meaningful;\footnote{We chose a correlation threshold of 
0.30 because it is a standard value for moderate correlations in social and urban sciences~\cite{cohen@1988}.}
\item \textit{Spearman} - computation of the Spearman's rank correlation coefficient over all aligned outliers; if the  coefficient is larger than 0.30 in modulus, 
and its p-value is lower than 0.05, the outlier relationship is considered meaningful.\footnote{Here, we chose a correlation 
threshold of 0.30 to be consistent with baseline \textit{Pearson}.}
\end{itemize}
}

%
%

%
While \tipo uses statistics to find meaningful alignments in an
unsupervised fashion, we also wanted to investigate how it compares to
learning-based approaches. 
To this end, we experimented with linear SVM~\cite{statistical-learning}
classifiers fitted over different sets of features. First, we used basic statistical features of attributes to verify
to what extent a simple summary  helps identify meaningful 
relationships across outliers. 
Next, we fitted the classifiers over features obtained
with \tipo (e.g., regression slopes) to understand whether they work best
as input for a machine learning model or as input for the statistical analysis that we propose in our method.
Concretely, these baselines were built as follows:
\begin{itemize}
\item \textit{SVM-Stats} - 
  given two attributes represented by dominant scores,
  the classifier uses as features the mean and standard deviation values of inliers,
  outliers, and all values.  We used the gold data to train the
  classifier and tested it with leave-one-out cross
  validation~\cite{statistical-learning}. The reported results are
  averages for all cross validation folds.
\item \textit{SVM-\tipo} - similar to \textit{SVM-Stats}, but the
  features used are derived from the execution of \tipo. They are:
  numbers of aligned values and outliers; $\bar{R}^2$; slopes,
  intercepts and their p-values; averages of error distributions for
  inlier, outliers, and all values.
\item \textit{SVM-Stats-synth} - similar to \textit{SVM-Stats} but,
  for each one of the 100 pairs in the gold data, we derived 10
  synthetic versions by adding to each attribute a Gaussian noise
  with mean of zero and standard deviation of 0.1. \tds{The goal was to introduce variability in a controlled manner
  for the purposes of bootstrapping additional data~\cite{EfroTibs93}.}\footnote{The
    proportion of true positives and true negatives in this synthetic
    data set differs from the original gold data  by less than
    5\%. This is evidence that the relationships detected in the gold
    data set were mostly preserved in the synthetic data, validating
    the use of the latter in our experiments.} We tested the
  classifier over the synthetic data with leave-one-out cross
  validation, and report averages over all cross
  validation folds. 
\item \textit{SVM-\tipo-synth}: it uses the same data generated for
  \textit{SVM-Stats-synth}, but using the features from \textit{SVM-\tipo}.  
\end{itemize}

%
\hide{Using \textit{SVM-Stats} and \textit{SVM-Stats-synth}, our goal is to
verify to what extent a simple summary of the attributes, consisting
of basic statistical features, helps identify meaningful relationships
across outliers. As for \textit{SVM-\tipo} and
\textit{SVM-\tipo-synth}, our goal is to understand whether features
obtained with \tipo, such as regression slopes and error
distributions, work best as input for a machine learning model or as
input for the statistical analysis that we propose in our method. \ab{Move this to before the list of synth datasets and reword accordingly. Otherwise it is not clear why these datasets are being generated. }}

The main challenge for these classifiers is that they require training
data, which is costly to generate because it requires human input. 
\tds{Moreover, note that outliers occur infrequently by definition and, due to that, having a large corpus of labeled outlier relationships is not realistic. 
Although these limitations suggest that classifiers are not appropriate for the 
analysis of outlier relationships, we still increased the training data artificially, as explained above, to verify whether the hypothetical use of a larger labeled corpus 
 could boost the classifiers' performances. We report results for classifiers trained on both gold and synthetic data.} 
%


Finally, we also compared \tipo against Data Polygamy
($DP$)~\cite{chirigati@sigmod2016}. DP identifies relationships across
numerical attributes using methods based on computational topology.
This comparison required extra steps such as additional human
annotation, and described in detail in
Section~\ref{subsec:dp-comparison}.
%

\paragraph{Clear Relationships. }
We start by discussing the results obtained for category \textit{Clear} because its examples have higher quality labels, as  \textit{all annotators} agreed on them. 
 As shown in Table~\ref{tab:baselines-clear}, \tipo outperforms all listed baselines in terms of F-measure.
In comparison with \textit{\tipo-Sub}, \tipo has significantly higher recall and comparable precision, suggesting that
using all aligned dominant scores instead of just dominant inliers leads to a better overall solution. 
Moreover, \tipo outperforms \textit{\tipo-MR} on all metrics, especially in terms of recall. This indicates that the incorporation of cumulative effects
is beneficial. 
The comparison between \tipo and \textit{\tipo-OLS} shows that the latter lacks balance: it has
a high precision  at the cost of a very low recall.
This happens because the aligned outliers are often not consistent with \trends computed with ordinary linear regressions. In fact, this parametrization ($\alpha = 1$)
  cancels the benefits of the \textit{Outlier-Biased} scheme, as it does not fit aligned outliers adequately. Consequently,  the \textit{clear positive} examples are not 
detected as such. This result illustrates the importance of the weighting scheme we propose: by fitting aligned outliers better, \tipo  detected 
most \textit{clear positive} examples. 
\tds{As for \textit{Pearson}, \tipo has lower recall but significantly higher precision because of its ability to prune false positives. In fact, 
while \textit{Pearson} detects 12 false positives, \tipo only detects 1. 
As mentioned above, \textit{Pearson}'s high false positive rate is likely a consequence of its low statistical power when applied over very small 
samples~\cite{sample-size,power-failure}. 
The results for \textit{Spearman} are more robust in contrast with \tipo, with comparable recall and F-measure. \textit{Spearman}'s precision, however, is 
considerably lower, also because of a more pronounced false positive rate (the baseline identified 4 false positives). 
These results suggest that these two baselines --- especially \textit{Spearman} --- could be good options for the detection of meaningful relationships over much 
larger samples of points, but it is hard to trust their statistical validity when there is little data available. Note that given the scarce nature of outliers 
and their alignments, it is not realistic to expect scenarios with large samples of aligned outliers, further reducing the potential usefulness of these baselines.} 
With respect to \textit{SVM-Stats}, \tipo outperformed it on all metrics. This suggests that basic statistical features, as expected, are not very informative of how 
meaningful the outlier relationships are. It is interesting to note, however, that this simple classifier performs substantially better than random guess, with 
recall similar to 
\textit{\tipo-Sub}'s and \textit{\tipo-MR}'s. 
\textit{SVM-\tipo}, on the other hand, outperformed \tipo on recall: it correctly identified one extra true positive. Despite that, the classifier's precision is 
significantly worse than that of \tipo. This suggests that the pruning of false positives, using features computed with \tipo, is better when a careful statistical analysis 
is performed, instead of when such features are the input to a classifier. 
Synthetic results obtained with \textit{SVM-Stats-synth} and \textit{SVM-\tipo-synth} reinforce the results of \textit{SVM-Stats} and \textit{SVM-\tipo} respectively.

\begin{table}[t!]
\centering
\begin{tabular}{llll}
\toprule
 &$R$ &$P$ &$FM$ \\
\midrule
\tipo & 0.88 & 0.96 & 0.92 \\ 
\textit{\tipo-Sub} & 0.64 & 1.00 & 0.78 \\
\textit{\tipo-MR} & 0.76 & 0.86 & 0.81  \\
\textit{\tipo-OLS} & 0.44 & 1.00 & 0.61  \\
\textit{Pearson} & 1.00 & 0.67 & 0.81  \\
\textit{Spearman} & 0.92 & 0.85 & 0.88\\
\textit{SVM-Stats} & 0.68 & 0.58 & 0.62 \\
\textit{SVM-\tipo} & 0.92 & 0.79 & 0.85 \\
\textit{SVM-Stats-synth}  & 0.61 & 0.79 & 0.68 \\
\textit{SVM-\tipo-synth}  & 0.86 & 0.75 & 0.80 \\
\bottomrule
\end{tabular}
\caption{Results generated by different metrics for category \textit{Clear}. $R$ stands for \textit{recall}; $P$ for 
 \textit{precision}; and $FM$ for \textit{F-measure}. The results for the SVM-based solutions are averages for all cross validation folds.}
\label{tab:baselines-clear}
\end{table}

\begin{figure}[t!]
    \centering
    \includegraphics[width=0.6\columnwidth]{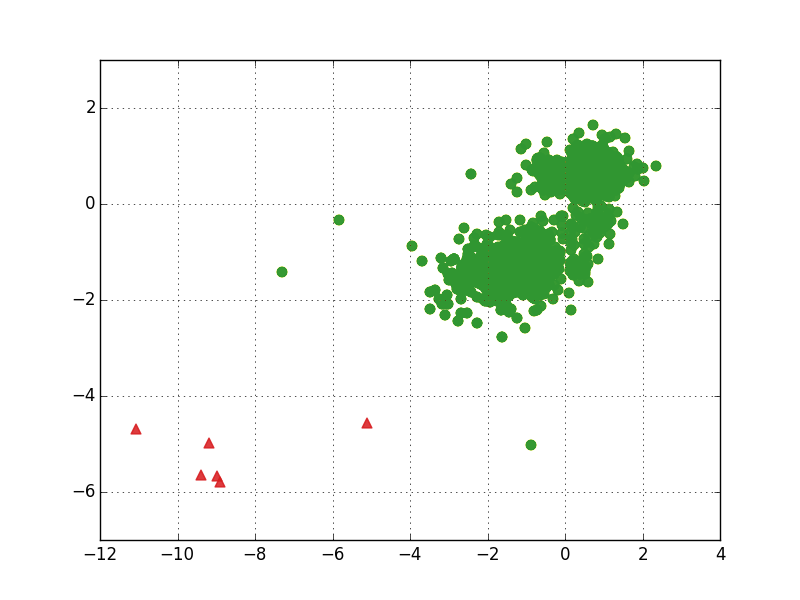}
\caption{A \textit{dubious negative} example that was labeled as \textit{negative} by 3 annotators and as \textit{positive} by 2. Both of its regression models ($f_1$ and 
$f_2$) pass the slope test and are reasonable, with $\bar{R}^2$ values above 0.80.}
\label{fig:dubious-negative}
\end{figure}

\paragraph{Dubious Relationships. }
The results for category \textit{Dubious} are summarized in Table~\ref{tab:baselines-dubious}.  
Note, however, that these results need to be taken with a grain of salt because the annotation quality of \textit{Dubious} is lower: users disagreed on the labels 
of all of its examples. In fact, we studied this category mostly to understand how \tipo behaves in scenarios where there is label 
uncertainty. 
The contrast between \tipo and \textit{\tipo-OLS} provides additional evidence that the \textit{Outlier-Biased} scheme is indeed effective for the detection of 
meaningful outlier relationships.
In comparison with \textit{\tipo-Sub}, \tipo has significantly higher recall and F-measure but lower precision. 
This happens because \tipo identified a larger number of false positives (10 out of 25, \textit{vs.} 1 out of 25 for \textit{Clear}). 
Specifically,  some  \textit{dubious negative} examples had reasonable regression models (see Definition~\ref{def:reasonable-goodness-of-fit}) with 
$\bar{R}^2$ values above 0.80, and their aligned outliers were consistent. \tipo then concluded that the aligned outliers of these examples 
 were meaningful---and, in fact, some annotators also came to this conclusion because they saw a trend-y pattern in these examples, but as for all 
 examples in the \textit{Dubious} category, there was no consensus as to whether the outlier relationships were meaningful. 
 Figure~\ref{fig:dubious-negative} shows a concrete example of a \textit{Dubious} false positive. 
\textit{\tipo-MR}, our best baseline,  performed slightly better than \tipo on all metrics. This happened because \textit{\tipo-MR} detected 17 true positives and 9 false
positives, while \tipo detected 16 true positives and 10 false positives. In the case of the true positive that was not detected by \tipo, the 
incorporation of cumulative effects increased the number of aligned outliers, and about 65\% of them were likely to be drawn from the model error distribution. 
Given that this value is slightly lower than $\beta = 0.67$, the example was not identified as positive. As for the false positive detected by \tipo but not by \textit{\tipo-MR}, we 
noticed that the model error distribution was again crucial: the number of aligned outliers bounded by this distribution was slightly higher than $\beta = 0.67$, 
leading to the extra false positive. 
These results suggest that the incorporation of cumulative effects, although mostly beneficial, may negatively affect the results, especially in cases 
 where the meaningfulness of the relationships is harder to evaluate (category \textit{Dubious}). 
\tds{As for baselines \textit{Pearson} and \textit{Spearman}, results show rather low values for all metrics. Not only do these methods identify a large number of 
false positives (11 out of 25 and 12 out of 25 respectively), but their ability to detect true positives is also lacking.  
These results may be a consequence of the fact that Pearson and Spearman's rank correlations do not handle the increased variability of small samples adequately, 
leading to unreliable correlation estimates.   
Moreover, these results reinforce the idea that such baselines are not suitable for the detection  of meaningful outlier relationships. }
With respect to the SVM-based classifiers, we have that both \textit{SVM-Stats} and \textit{SVM-\tipo} have good recall when compared to the other solutions. 
In fact, the former detected 17 true positives just like \textit{\tipo-MR}, outperforming \tipo on recall as well. Regardless, their significantly lower precision leads to 
 worse \textit{F-measure}, suggesting that \tipo is better suited for the task. 
The results for \textit{SVM-Stats-synth} and \textit{SVM-\tipo-synth} are also worse than \tipo's overall, but they show more balance between precision and recall. 
In this case, more data helped SVM make better predictions. Note, however, that \textit{SVM-Stats-synth} and \textit{SVM-\tipo-synth} were not run on the original gold data, 
so comparisons between these classifiers and \tipo are limited by design.  

\begin{table}[t!]
\centering
\begin{tabular}{llll}
\toprule
 &$R$ &$P$ &$FM$ \\
\midrule
\tipo & 0.64 & 0.62 & 0.63 \\ 
\textit{\tipo-Sub} & 0.32 & 0.89 & 0.47 \\
\textit{\tipo-MR} & 0.68 & 0.65 & 0.67  \\
\textit{\tipo-OLS} & 0.44 & 1.00 & 0.61  \\
\textit{Pearson} & 0.32 & 0.42 & 0.36  \\
\textit{Spearman} & 0.48 & 0.50 & 0.49 \\
\textit{SVM-Stats} & 0.68 & 0.49 & 0.57 \\
\textit{SVM-\tipo} & 0.64 & 0.52 & 0.57 \\
\textit{SVM-Stats-synth}  & 0.59 & 0.62 & 0.60 \\
\textit{SVM-\tipo-synth}  & 0.60 & 0.58 & 0.59\\
\bottomrule
\end{tabular}
\caption{Results generated by different metrics for category \textit{Dubious}. $R$ stands for \textit{recall}; $P$ for 
 \textit{precision}; and $FM$ for \textit{F-measure}.}
\label{tab:baselines-dubious}
\end{table}

\begin{figure}[t!]
    \centering
    \includegraphics[width=0.6\columnwidth]{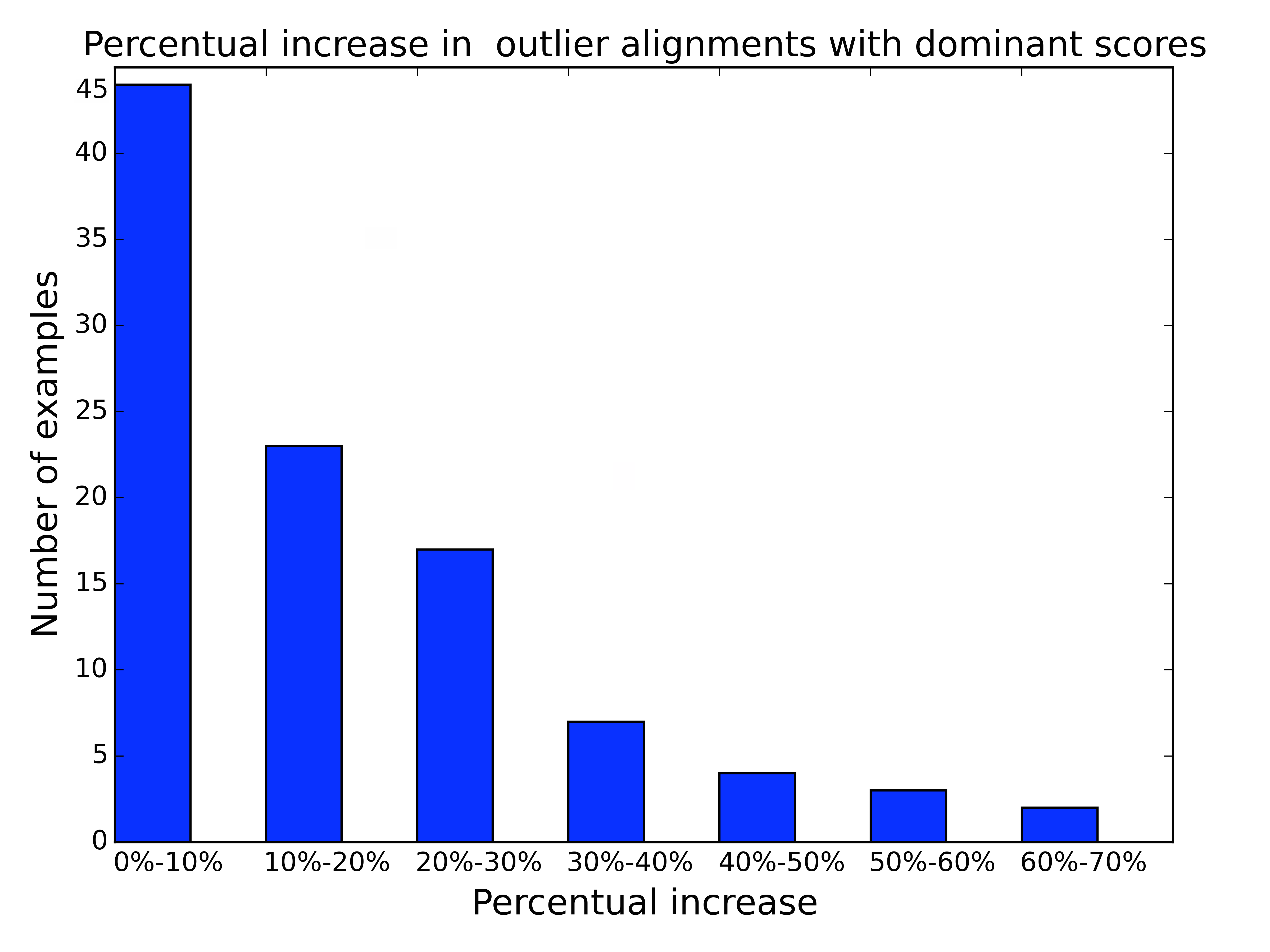}
\caption{The use of dominant scores increased the number of outlier alignments for 56 out of the 100 gold data examples. On average, the number of outlier alignments increased in 
 about 15\% of instances. \hide{Is ``percentual'' a legit word? Percentage could be more appropriate. }}
\label{fig:increase-outlier-alignments}
\end{figure}

\paragraph{Discussion.}
Overall, these quantitative results suggest that the detection of
meaningful outlier relationships is a hard problem.  They also show
that \tipo is effective, providing a trade-off between true and false
positives that is consistently competitive.
The use of cumulative effects increased the number of outlier
alignments for most cases, as shown in
Figure~\ref{fig:increase-outlier-alignments}. This factor, along with
the \textit{Outlier-Biased} scheme, specifically contribute to the
recall of our solution. Besides that, we noted that component
\textit{Meaningfulness Verification}, by testing the quality of models
and the consistency of aligned outliers, significantly contributes to
the precision of \tipo.
These results also indicate that verifying \textit{statistical
  meaningfulness} is consistent with, for the most part, the annotators' idea of
meaningfulness.

\tipo, as illustrated by these experiments, is particularly useful to
further understand whether outlier alignments are likely to be
coincidental or, rather, help explain each other. To the best of our
knowledge, there is no previous work that specifically addresses 
 a formalization of the concept of meaningfulness that can be
backed up statistically and matches users' intuition. Our solution,
however, has certain limitations: the notion of meaningfulness is tied
to linear relationships; attributes have to be numerical; outliers
need to be large (in modulus); and one needs to
calibrate a few parameters. In the future, our goal is to address all
these restrictions.

 
\subsection{Effectiveness: Comparison with Data Polygamy}
\label{subsec:dp-comparison}

Data Polygamy ($DP$)~\cite{chirigati@sigmod2016} is a framework designed to detect 
statistically significant relationships between attributes of (spatio-)temporal data sets. 
To the best of our knowledge, $DP$ is the published work that is closest to ours in terms of the problem it addresses. 
Given a pair of attributes as input, $DP$ models their individual values as 
scalar functions, which provide a mathematical representation of their \textit{topological terrain}. 
The framework then derives a relationship for a pair of scalar
functions when their \textit{topological features}, i.e., peaks and
valleys, overlap. 
$DP$ generates two types of topological features: \textit{salient} and \textit{extreme}. The former corresponds to data values 
that are moderately different from their (spatio-)temporal vicinity, albeit still quite common in the data; the latter corresponds to 
outliers among salient features which occur very rarely. We thus focus
our comparison on the latter. 
%

\paragraph{Choice of input data for $DP$. } We started off by running $DP$ over all pairs of original attributes
from the data sets in Table~\ref{tab:data-sets}, but the number of
extreme features per attribute was fairly small: around three
  extreme features on average were identified for each original
  attribute. This resulted in very few overlaps across attributes and no
statistically significant extreme relationship. 
To increase the number of identified extreme features, we then ran $DP$ 
over pairs of mean residual representations instead. 
Mean residual representations, computed with different time window sizes~\footnote{
The time window sizes that we used were of 28, 30, 64, 90, 128, 180, 256, 360, and 365 days,  as 
indicated in Section~\ref{subsec:experimental-setup}.}, uncover a range of 
local and global outliers that may not be as visible in the original attributes. Our intuition then was that  
$DP$ would identify more extreme features, and potentially more feature overlaps, when executed over pairs of mean residual representations. 
More extreme features (on average, eight per representation) and extreme feature overlaps were indeed identified, but again   
no extreme relationships were considered statistically significant. 
In fact, we noticed that $DP$ filters out relationships
across small samples of extreme features because of the way it tests
for statistical significance.

\begin{figure}[t!]
    \centering
    \subfloat[\textit{clear positive}]{\includegraphics[width=0.35\textwidth]{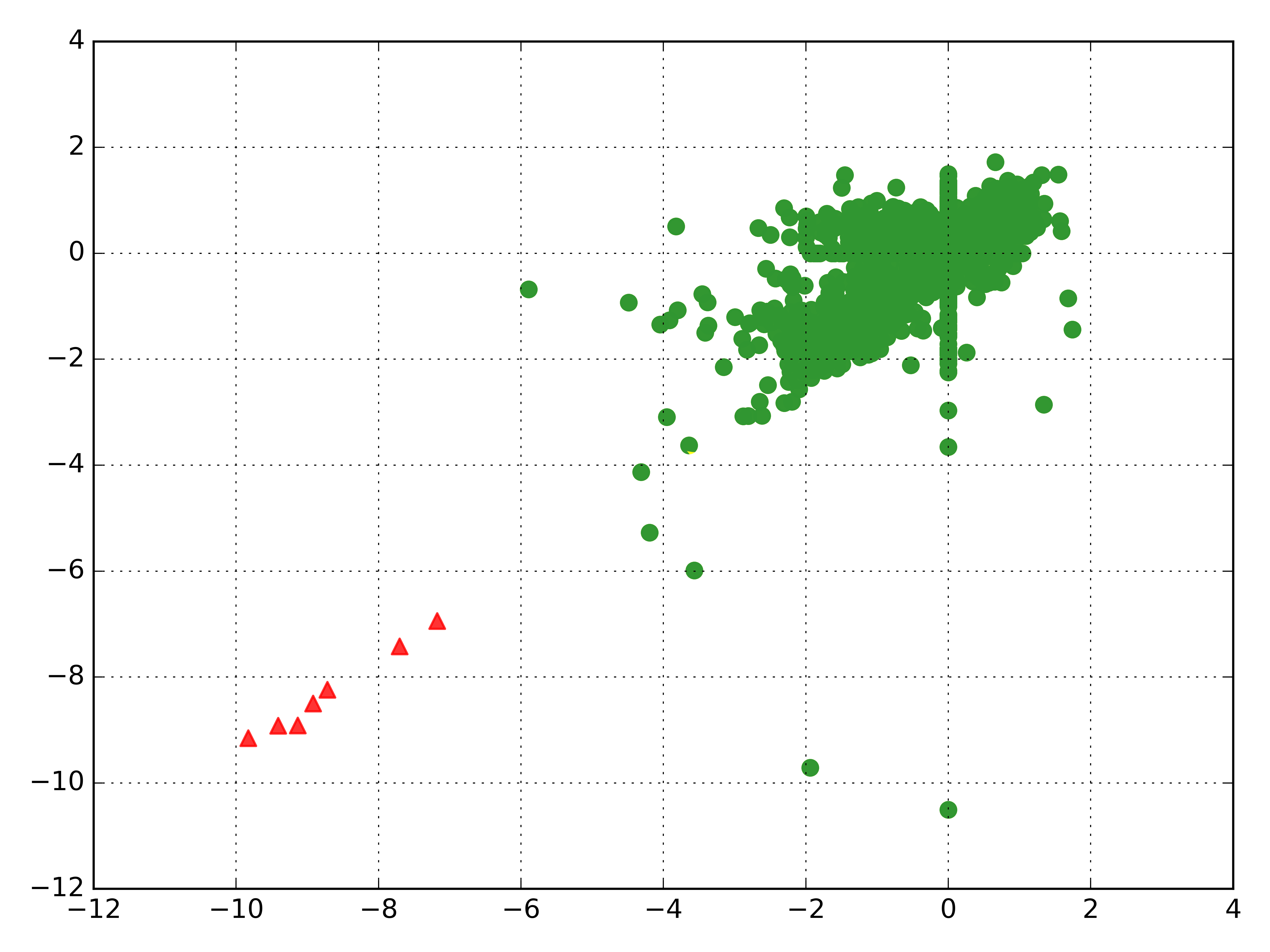} \label{subfig:dp_easy_positive}}
    \qquad
    \subfloat[\textit{clear negative}]{\includegraphics[width=0.35\textwidth]{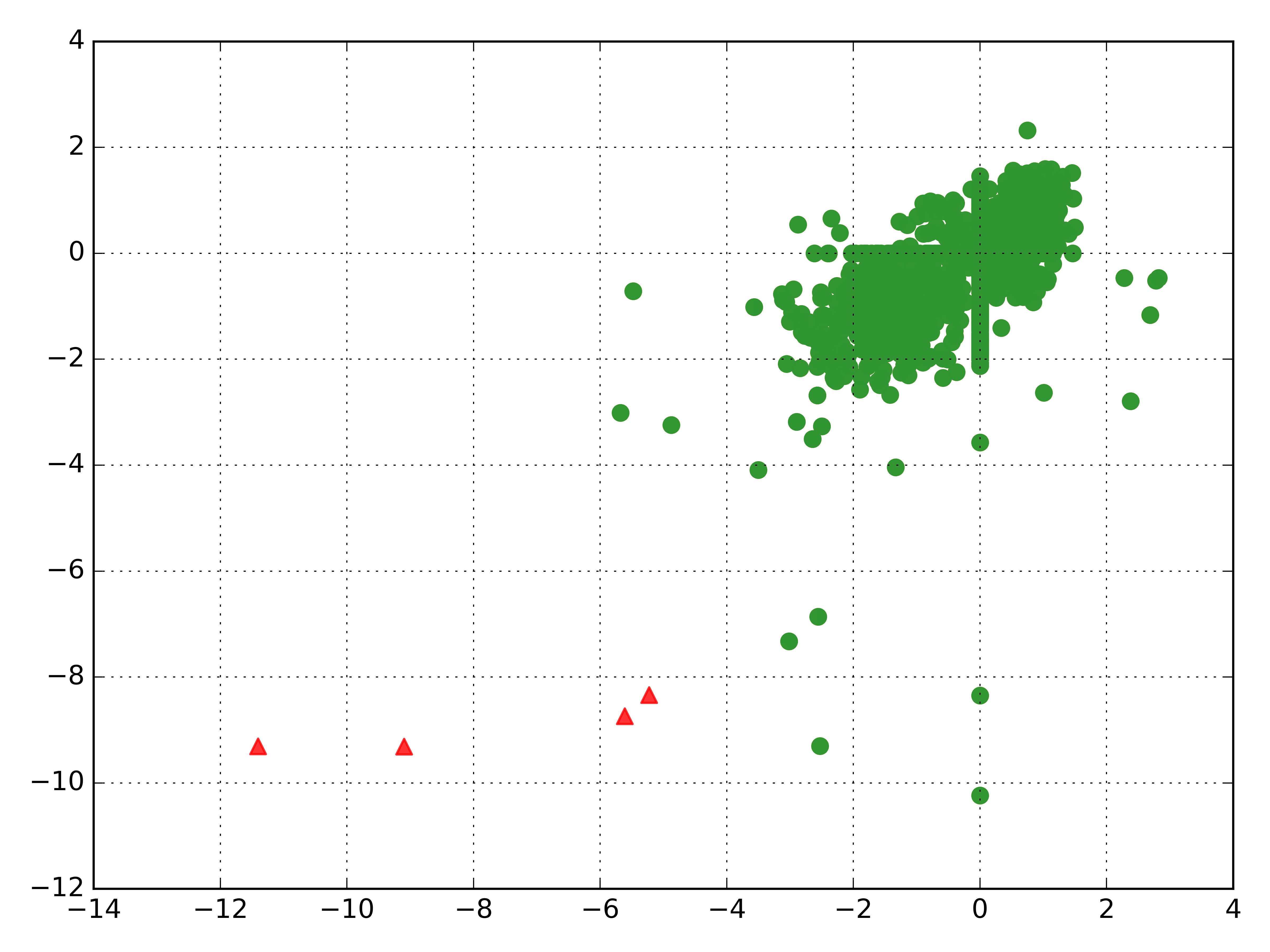} \label{subfig:dp_easy_negative}}
\caption{Examples for categories \textit{clear positive} and \textit{clear negative}. The values in the scatterplot are co-occurring scalar function values, generated by $DP$.}
\label{fig:dp-gold-data-examples}
\end{figure}

Given that $DP$ did not consider any extreme relationship significant,
we wanted to understand whether \tipo detects any of them as
meaningful, and if its results are compatible with users'
expectations.  
To this end, we generated scatterplots like the ones in Figures~\ref{subfig:dp_easy_positive}~and~\ref{subfig:dp_easy_negative} for all
pairs of \textit{co-occurring scalar function values} generated by $DP$, as long as the pairs had at least one extreme feature overlap. 
%
 The red triangles 
correspond to overlapping extreme features, i.e., extreme features 
that co-occurred on a same day, and the green circles represent  other co-occurring scalar function values. 
The scatterplots were labeled by five 
different annotators, following a procedure similar to the one described in Section~\ref{subsec:gold-data}. We asked each annotator the following question: 
\textit{Do you see a meaningful pattern across the aligned outliers?}, indicating that they corresponded to the red triangles. 
Annotators again interpreted the term \textit{meaningful} in slightly
different ways, but focused on a few aspects: the presence of clusters
of red triangles, the existence of a trend pattern across them, a
certain shape coherence with close green circles, and the presence of
at least two red triangles. For each scatterplot, the corresponding
pair of representations was annotated as \textit{clear positive},
\textit{clear negative}, \textit{dubious positive}, or \textit{dubious
  negative}, following the same semantics described in
Section~\ref{subsec:gold-data}. Due to the small number of examples,
as there were not many co-occurring extreme features, we focused on
the \textit{clear positive} and \textit{clear negative} categories,
gathering 10 examples of
each. Figures~\ref{subfig:dp_easy_positive}~and~\ref{subfig:dp_easy_negative}
are scatterplots illustrating these two categories.

After annotating the $DP$-based scatterplots, we ran \tipo over their
corresponding attribute pairs to verify if it is compatible with the
human labeling.
It is important to mention that the scalar functions used in $DP$ satisfy the requirement for outlier detection functions 
 $\psi_X$. Consequently, it was straightforward to run \tipo over the annotated data.  
In this experiment, we obtained a recall of 0.90, a precision of 0.64, and an F-measure of 0.75.
These results indicate that our approach is useful for capturing patterns across aligned outliers that seem meaningful to users.
Still, note that \tipo's precision is significantly lower than its
recall. This happened because our method detected examples with a
single outlier co-occurrence as meaningful, specifically because the
co-occurrence in question was consistent with the computed regressions
(see Definition~\ref{def:consistency}). The annotators, on the other
hand, did not consider that any meaningful pattern could be drawn from a
single outlier co-occurrence.
In practice, however, it is possible to have a single outlier co-occurrence that actually uncovers a meaningful relationship: for example, a hurricane in a big city 
may happen only once, but it does help explain concomitant power outages.
Annotators probably did not consider such cases because the semantics of the data (e.g., axes indicating the corresponding original attributes) 
was not presented to them: without any further knowledge of the data, they relied on a more generic intuition, ruling out relationships 
involving single outlier co-occurrences. 
Our conclusion is that hiding the semantics of the data prevented annotation biases, but it also limited our experiment. After all,   
semantics, just like statistical intuition, plays an important role in understanding whether an outlier relationship is meaningful.

This experiment indicates that $DP$ is competitive for the detection of significant relationships across \textit{salient features}, which are relatively common in 
the studied data sets.
However, this does not seem to be the case for relationships across small samples of outliers (\textit{extreme features}). 
In fact, $DP$ did not identify any of the annotated examples as statistically significant, strongly suggesting that \tipo is more suitable for the task of detecting important outlier relationships. 
Although the experiments presented in this section are fairly limited, using small amounts of annotated data, it seems that \tipo is capable of
providing insights that are useful  for outlier relationships, whereas $DP$ simply filters them out. We believe that both techniques are 
mainly suited for distinct, albeit strongly related, applications, and their outputs complement
one another in the exploration of temporal data. 

\subsection{Effectiveness: Case Studies}
\label{subsec:case-studies}

While the experiments in
Sections~\ref{subsec:effectiveness-baselines}~and~\ref{subsec:dp-comparison}
provide quantitative evidence of the effectiveness of \tipo, in what
follows, we use real use cases to assess its usefulness in practice, i.e., if the meaningful outlier 
relationships it detects are interesting and seem useful. 

\begin{figure}[tb]
    \centering
    \subfloat[Crime and temperature]{\includegraphics[width=0.32\columnwidth]{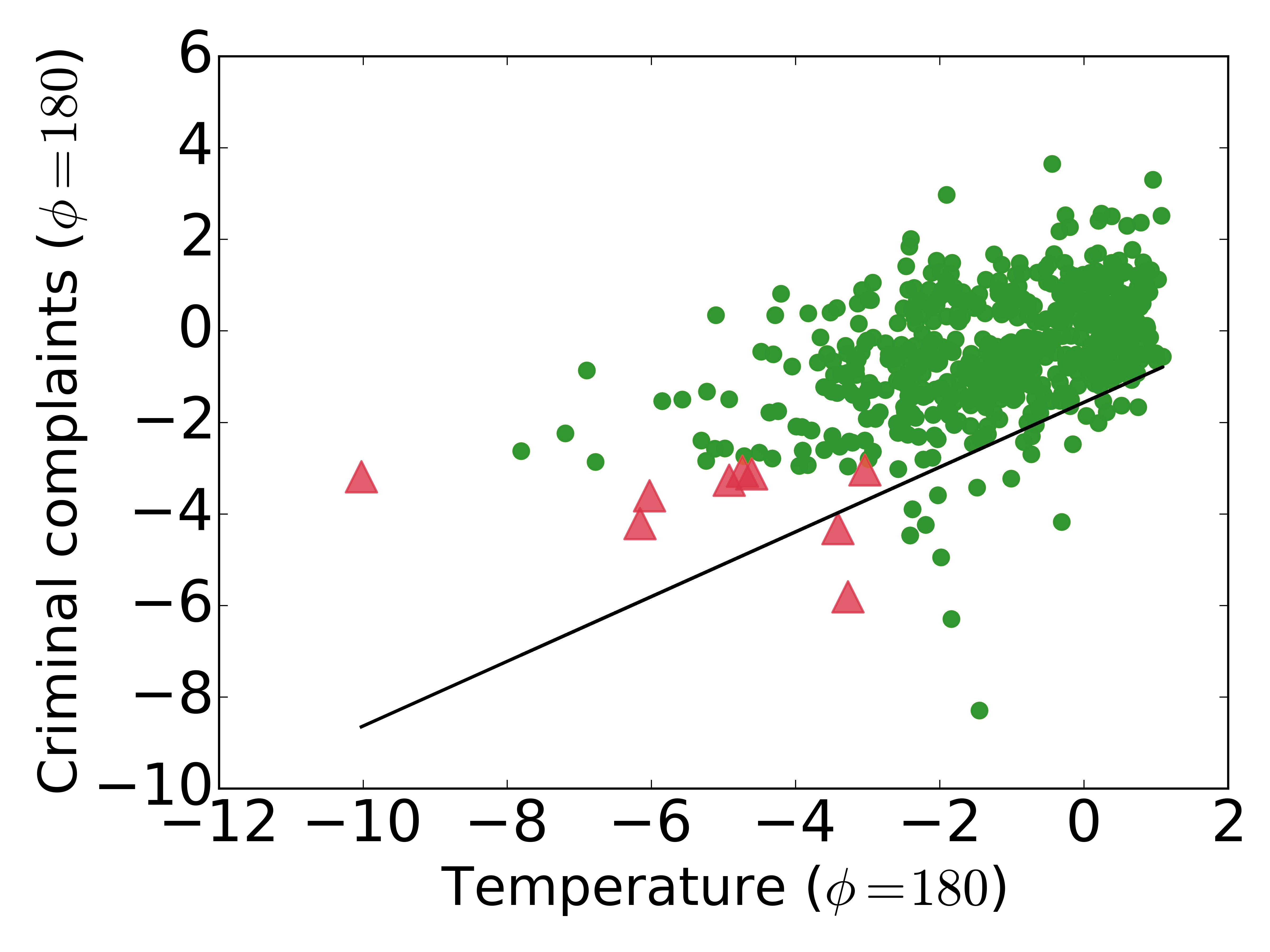} 
   \label{case:180-complaints-180-temperature}}
    \subfloat[Collisions and taxi trips]{\includegraphics[width=0.32\columnwidth]{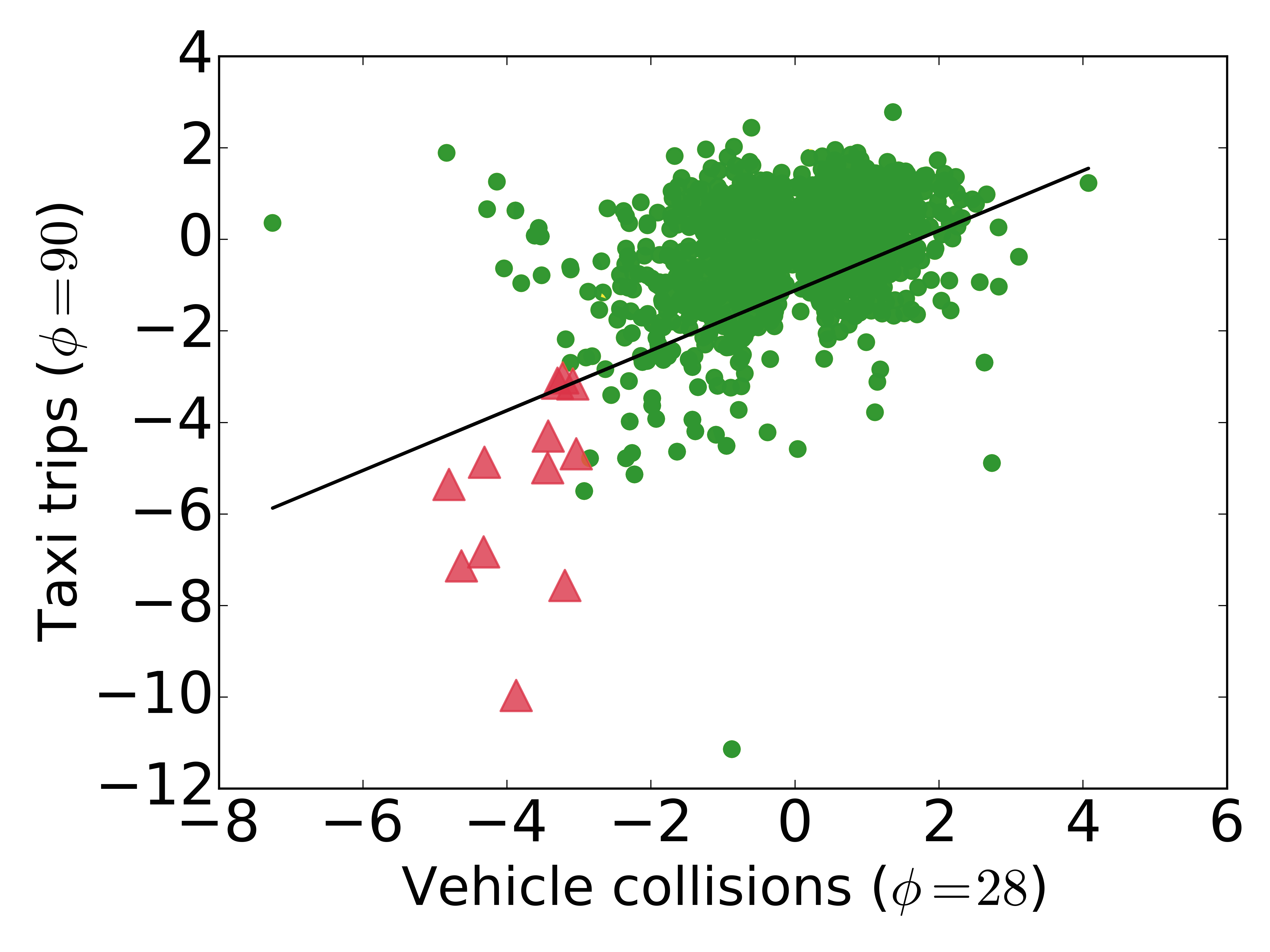} 
    \label{case:28-crash-90-taxi}}
     \subfloat[Temperature and restaurant complaints]{\includegraphics[width=0.32\columnwidth]{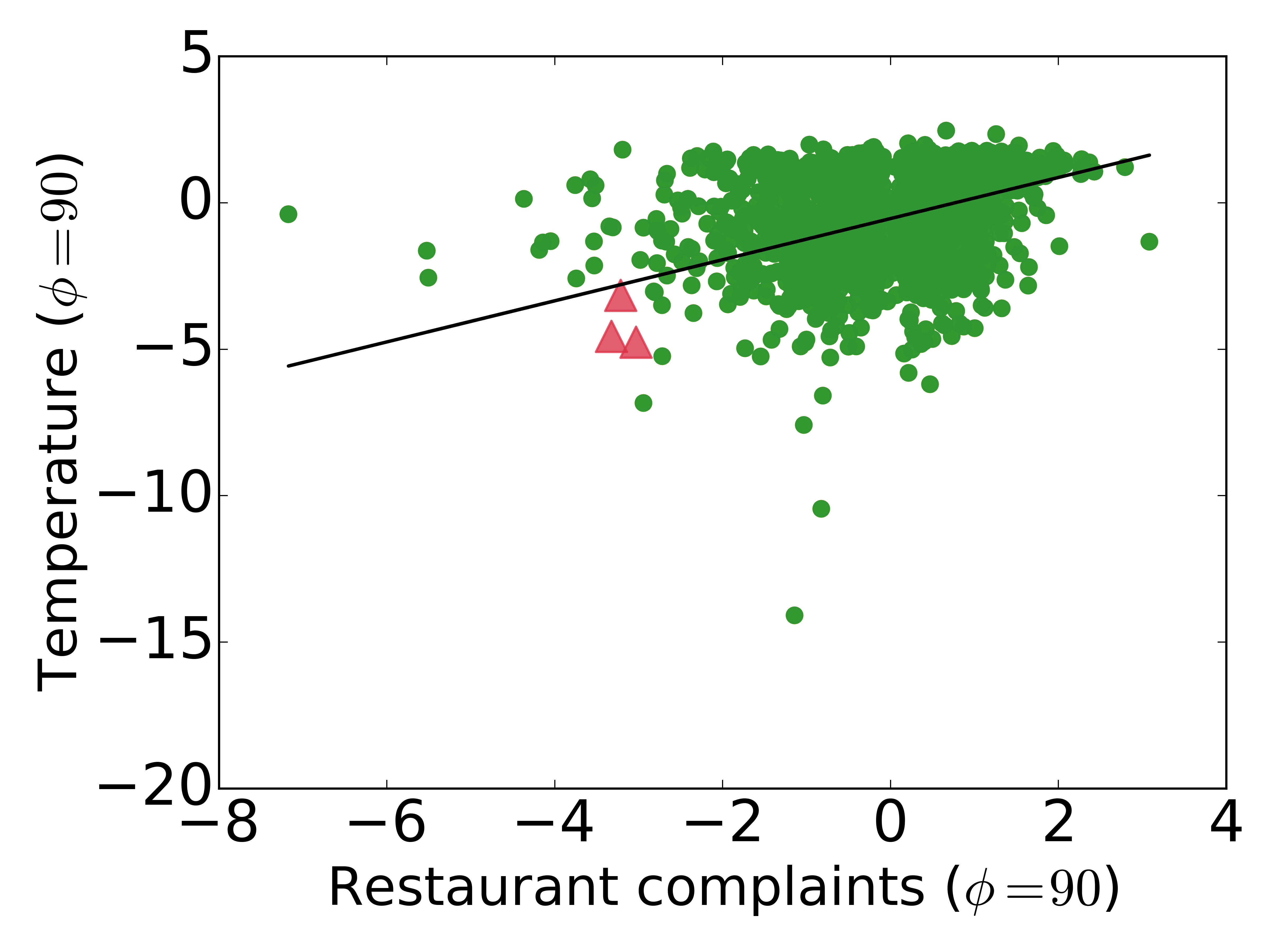} 
     \label{case:90-temperature-90-restaurant-complaint}}
\caption{Scatterplots showing outlier relationships detected as meaningful for different attribute combinations. The attribute values are represented with dominant scores 
built over mean residuals with  different time window sizes $\phi$. Black lines correspond to $WLS$ models.}
\label{fig:cases}
\end{figure}

\paragraph{Criminal Offenses and Temperature.} 
%
While published studies reported links between high temperatures and 
increases in criminal activity~\cite{chinasi@jurbanhealth}, we obtained a new insight through a
relationship discovered by \tipo: low values for
temperature are aligned with low values for offenses. 
Consider Figure~\ref{case:180-complaints-180-temperature}.
In the positive
quadrant of the plot ($x > 0$ and $y>0$), there is indeed a positive
\trend between the two attribute representations, but the alignments
are not as extreme as in the negative quadrant. In other words, the
explanation of the outliers -- in this case, very safe and cold days,
complements the prior knowledge that hot days are linked to higher
crime. 
This underscores the utility of techniques like the one we propose,
which can be applied in addition to traditional statistical techniques
for finding correlations.
This finding can have implications for decisions regarding safety investments, made by the
NYC administration. \hide{New York City makes
substantial investments and increases police presence to ensure safety
for the New Year's Eve celebrations~\cite{amny-bombs}.  However, New
Year's Eve celebrations often occur on very cold days when few crimes
are committed.} 
As we discussed in Section~\ref{sec:intro}, while our method
identifies outlier \trends that can lead to new hypotheses (e.g., in
this case the association between low temperature and low crime),
domain experts must further explore the data to prove (or disprove)
the hypotheses. In this example, through further investigation, an
expert found that in most years, temperatures on December 31st are
low. One exception happened on December 31, 2011: a comparatively warm
day, with a low temperature of 8C and a significantly higher number of
criminal offenses.
The relationship \tipo discovered followed by additional investigation
could help the NYPD construct predictive models that enable them to
adjust their policing investments thus leading to cost savings without
jeopardizing safety.

\paragraph{Vehicle Collisions and Taxi Trips.}
We identified an extreme relationship involving abnormally small
numbers of vehicle collisions ($\phi = 90$) and taxi trips
($\phi = 28$), shown in Figure~\ref{case:28-crash-90-taxi}.
The most extreme co-occurrence corresponds to October 29, 2012, during
Hurricane Sandy~\cite{wikipedia-sandy}. However, a pattern can already
be seen across less extreme co-occurrences.
Two of these co-occurrences 
correspond to 2013's Independence Day and Memorial Day holidays.  Our
hypothesis is that the number of taxi trips is a proxy for the total
number of cars in New York City, and while the number of trips does
not help understand most variations in the number of collisions,
including its peaks, its valleys are a consequence of a quieter
period, when many people leave for the holidays.

\paragraph{Restaurant Complaints and Temperature.} An interesting
relationship detected by \tipo, shown in
Figure~\ref{case:90-temperature-90-restaurant-complaint}, involves
\textit{seasonal} negative temperature outliers ($\phi = 90$) and \textit{seasonal} 
low numbers of 311 restaurant complaints ($\phi = 90$).
By checking complaint descriptions in the 311 data set, we found that
most restaurant complaints have to do with: the presence of pests (e.g., rodents) and
garbage in the food preparation area; spoiled food; lack of letter
grading;\footnote{In New York City, the Health Department conducts
  unannounced inspections of restaurants at least once a year. At the
  end of the inspection, the inspector gives a grade to the restaurant
  --- the fewer violations found, the higher the
  grade~\cite{restaurant-inspection}.} and foreign objects in the
food.
Although it makes sense to expect that relatively lower
temperatures may attract fewer pests, or even contribute to reduced
food spoilage, it is surprising that complaints that seem
season-independent, such as letter grading or food contamination, also 
become infrequent when the temperature suddently drops.
In fact, there is evidence that the number of customers in restaurants is 
relatively constant throughout the seasons---except during Summer, when
establishments tend to be more empty~\cite{nyc-summer-eat-out}---,
and at first we thought that this implied that the number of complaints should also be relatively constant. 
%
%
%
Despite this evidence, however, some relatively low numbers of complaints are definitely linked to sudden decreases in temperature in the data we studied,
and our hypotheses are that (1) either customers complain less on such days because they also go out less (i.e., restaurant attendance is not that constant after all), or maybe spend less time
in the establishments, or (2) this association is a data quality issue. 
%
%
%
This case study is thus surprising precisely because it needs more domain knowledge, or data sources, to be better understood. 
It raised the question of whether it actually reflects 
a systematic relationship between restaurant complaints and temperature, or
whether it corresponds to a relevant data quality issue. An expert on these
data sets could use this finding to help build an informed analysis on the connection
between seasonally low temperatures and restaurant complaints, but this is outside the scope of our work.


\subsection{Scalability Evaluation}
\label{subsec:scalability}

\hide{complaint: D3. In Table 4, the time cost of the tested method largely increase with the increase of sample size. For example, when the sample size increase from 100 to 200 (twice size), time cost increases from 341 min to 1402 min (about 4 times). The authors should explain more on the reason, which verifies the scalability of the proposed methods. // every new dataset we introdce has the same probability of matching. I did not make it in such a way that, as I add more
data, I also decrease the prob. of matching.
This is more ``index-y''... The current speedup suggests that the use of the index prunes away 1/3 of the pairs. make an experiment where the addition of data is more like
what happens in traditional indexing scenarios: you add more data but the match is very selective. you don't keep adding stuff and the matching probability remains the same.
one type of experiment: something grows linearly in the size of the data. (the case here)canother kind: the number of matches is constant but the data keeps growing. maybe
you have a log kind of behavior but it doesn't grow linearly. maybe report the experiments for a single dataset (a target dataset). clarify the text in this section as well:
it seems like the reviewer understood something a bit different.}

\begin{table*}[t]
\centering
\small
\begin{tabular}{p{1.5cm}p{1.5cm}p{1.5cm}p{1.75cm}p{1.75cm}p{1.75cm}p{1.5cm}}
\toprule
\textbf{Sample Size} & \textbf{Total Pairs} & \textbf{Indexed Pairs} & \textbf{Fraction of Pruned Pairs} & \textbf{Time w/o index (\textit{min})}  & \textbf{Time w/ index (\textit{min})}  & \textbf{Index Speedup} \\ 
10 & 45 & 32 & 0.29 & 3.32 & 2.55 & 1.30 \\
\midrule
25 & 300 & 209 & 0.30 & 23.96 & 17.01 & 1.41  \\
\midrule
50 & 1225 & 798 & 0.35 &  90.33 & 59.12 & 1.53 \\ 
\midrule
100 & 4950 & 2966 & 0.40 &  341.24 & 216.13 & 1.58 \\
\midrule
200 & 19900 & 12582 & 0.37 & 1402.75 & 869.19 & 1.61 \\
\bottomrule
\end{tabular}
\caption{Scalability results. The fraction of pruned attribute pairs is roughly one third for every sample size.}
\label{tab:scalability}
\end{table*}

To evaluate the scalability of \tipo, we performed two experiments that measure how  its execution time increases as the number
of attributes grows. The experiments were carried out  on a desktop with an Intel(R) Xeon(R) Processor E5-2630 
(4x8 cores) running at 2.40GHz, with 32GB of RAM. \tipo is implemented in Python 2.7. 

In the first experiment, we simulate a scenario where the fraction of pruned attribute pairs is roughly the same for differently-sized collections.  
First, we generated differently-sized random samples of attributes from our collection (Table~\ref{tab:data-sets}) and represented them
with mean residuals. Next, we compared the execution time of \tipo 
over all distinct pairs in the samples with and without the alignment 
index.\footnote{Note that, as we experiment with 9 different time
  window sizes $\phi$ and there are 84 attributes in
  Table~\ref{tab:data-sets}, there is a total of 756 mean residual
  representations for the attributes.}
The results shown in Table~\ref{tab:scalability} correspond to the
median value among 5 cold-cache runs for executions with and without
the index.
\tds{Note that the execution times grow roughly linearly on the total number of attribute pairs, both with and without the index. When the index is used, however, the execution 
times consistently decrease by about a third ---
a consequence of pruning around one third of the  attribute pairs for every sample size. These results indicate that, although the relationship between the total number of attribute pairs and the execution times does not become sub-linear with the index, the speedup has a significant practical impact, especially for larger attribute samples.}
\tds{In the next experiment, we simulated a different scenario: as the number of attributes in a collection grows, the fraction of attribute pairs with aligned outliers 
decreases significantly. 
%
%
In such cases, our alignment index should have a more substantial impact on the execution time of \tipo.
%
The motivation behind this experiment comes from traditional indexing scenarios, where the probability of having two items in the same posting
list\footnote{Recall that in the alignment index we propose, posting lists are composed of attributes with aligned outliers.} decreases as more data is added to the index~\cite{baeza2011IR}.
To run this experiment, we first represented attributes from our  collection with mean residuals, and then generated differently-sized attribute samples with the same 
number of indexable attribute pairs (i.e., pairs with aligned outliers). By keeping the number of indexable pairs constant, we varied the sample size (and, consequently,  the 
total number of attribute pairs) to simulate different levels of index pruning. Next, we compared the execution time of \tipo over all samples  with and without the 
alignment index.
As before, results in Table~\ref{tab:scalability2} correspond to the median value among 5 cold-cache runs. 
%
%
Note that the execution time remains almost constant when the index is used --- when it is not, the execution time grows roughly linearly on the total 
number of attributes. 
As a consequence, the index speedup significantly increases as the sample size grows. 
This suggests that the alignment index is particularly beneficial for large collections with a small fraction of indexable pairs, which may be common in practice.}

%


\begin{table*}[t]
\centering
\small
\begin{tabular}{p{1.5cm}p{1.5cm}p{1.5cm}p{1.75cm}p{1.75cm}p{1.75cm}p{1.5cm}}
\toprule
\textbf{Sample Size} & \textbf{Total Pairs} & \textbf{Indexed Pairs} & \textbf{Fraction of Pruned Pairs} & \textbf{Time w/o index (\textit{min})}  & \textbf{Time w/ index (\textit{min})}  & \textbf{Index Speedup} \\ 
25 & 300 & 209 & 0.30 & 23.18 & 16.29 & 1.42 \\
\midrule
30 & 435 & 209 & 0.52 & 32.54 & 16.42 &  1.98 \\
\midrule
40 & 780 & 209 & 0.73 &  59.88 & 16.63  & 3.60 \\
\midrule
70 & 2415 & 209 & 0.91 & 186.13 & 16.89 &  11.02 \\
\midrule
\bottomrule
\end{tabular}
\caption{\tds{Scalability results. The number of indexed pairs remains constant for different sample sizes.}}
\label{tab:scalability2}
\end{table*}

Both experiments demonstrate the importance of using an index in distinct scenarios, and indicate that \tipo is scalable and can be applied
over large collections of data sets.
Note that we ran the experiments sequentially, on a desktop. However,
the computation is easily parallelizable: each pair can be computed
independently. Therefore, the times reported in
Tables~\ref{tab:scalability}~and~\ref{tab:scalability2} would be greatly reduced if the
experiments were run in parallel on a cluster.


\subsection{Sensitivity to Parameter Variations}
\label{subsec:alpha-beta-variations}

In what follows, we assess the robustness of \tipo with respect to slight variations in its parameters. The discussion is based on our experiments, the data, and 
the annotation scheme we used in Section~\ref{subsec:gold-data}. Results for gold data categories \textit{Clear} and \textit{Dubious} were generated separately. 

\begin{figure}[t!]
    \centering
    \subfloat[]{\includegraphics[width=0.45\textwidth]{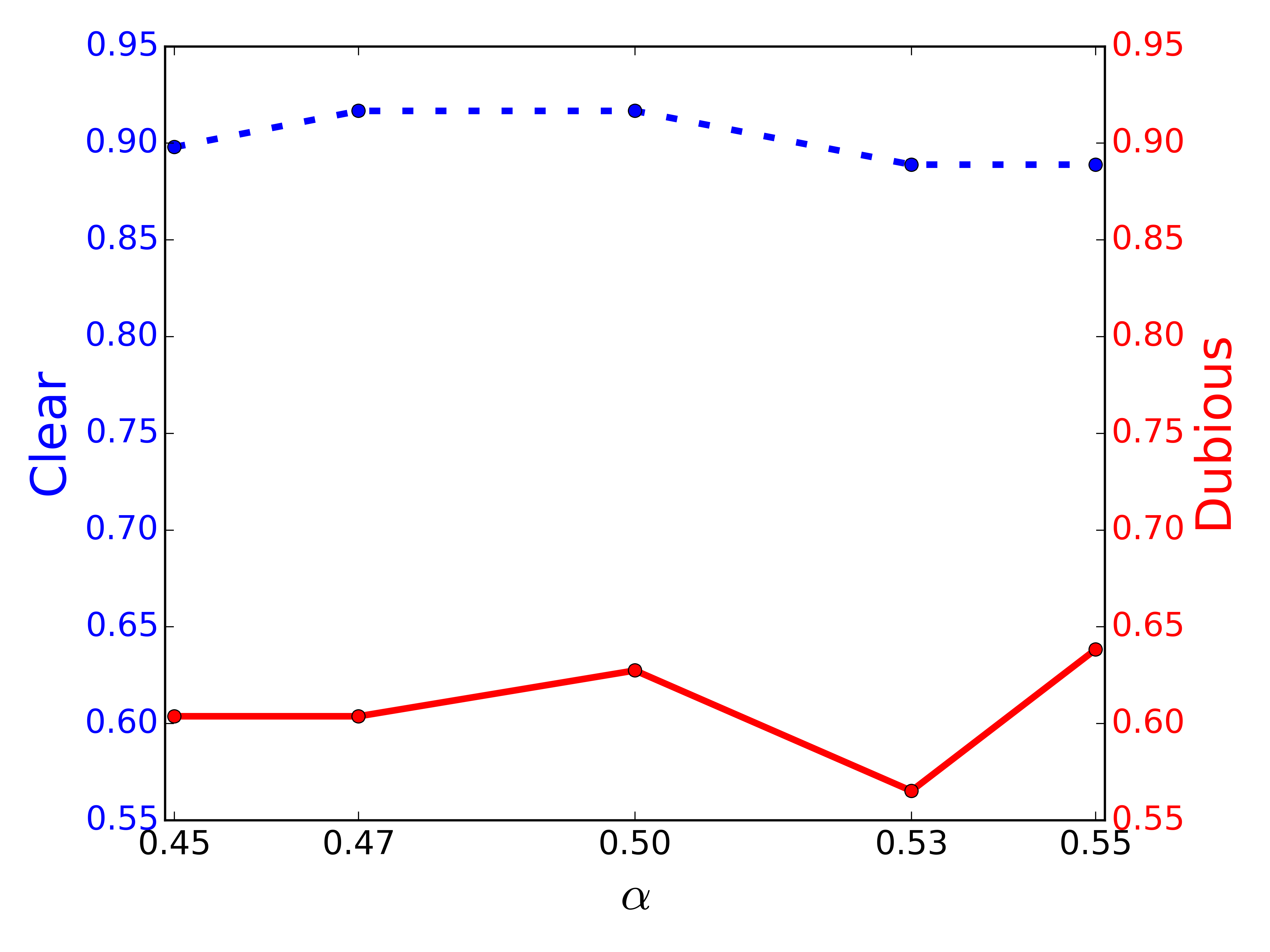} 
    \label{fig:robust-alpha}}
    \qquad
    \subfloat[]{\includegraphics[width=0.45\textwidth]{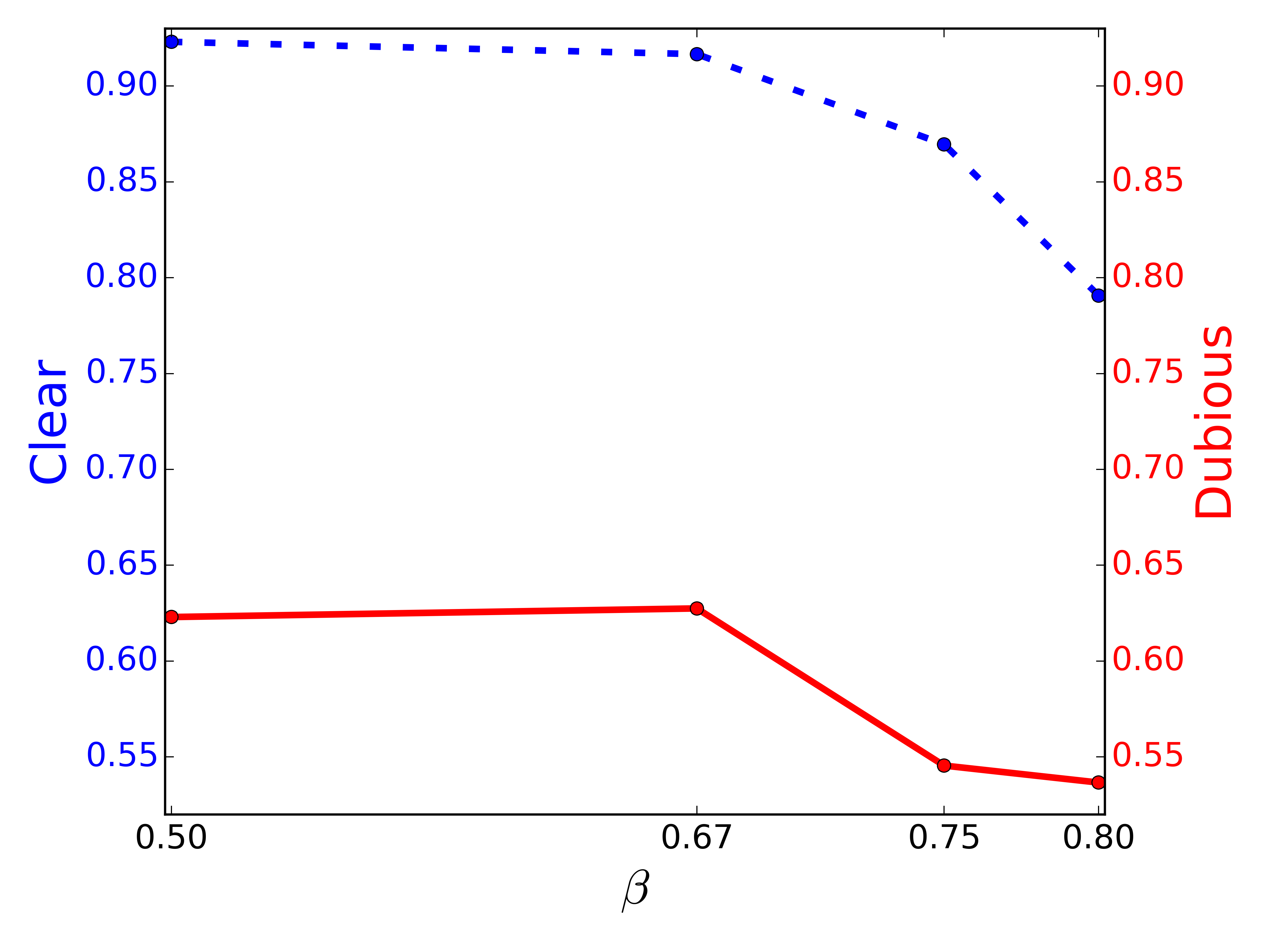} 
    \label{fig:beta-variation}}
    \qquad
    \subfloat[]{\includegraphics[width=0.45\textwidth]{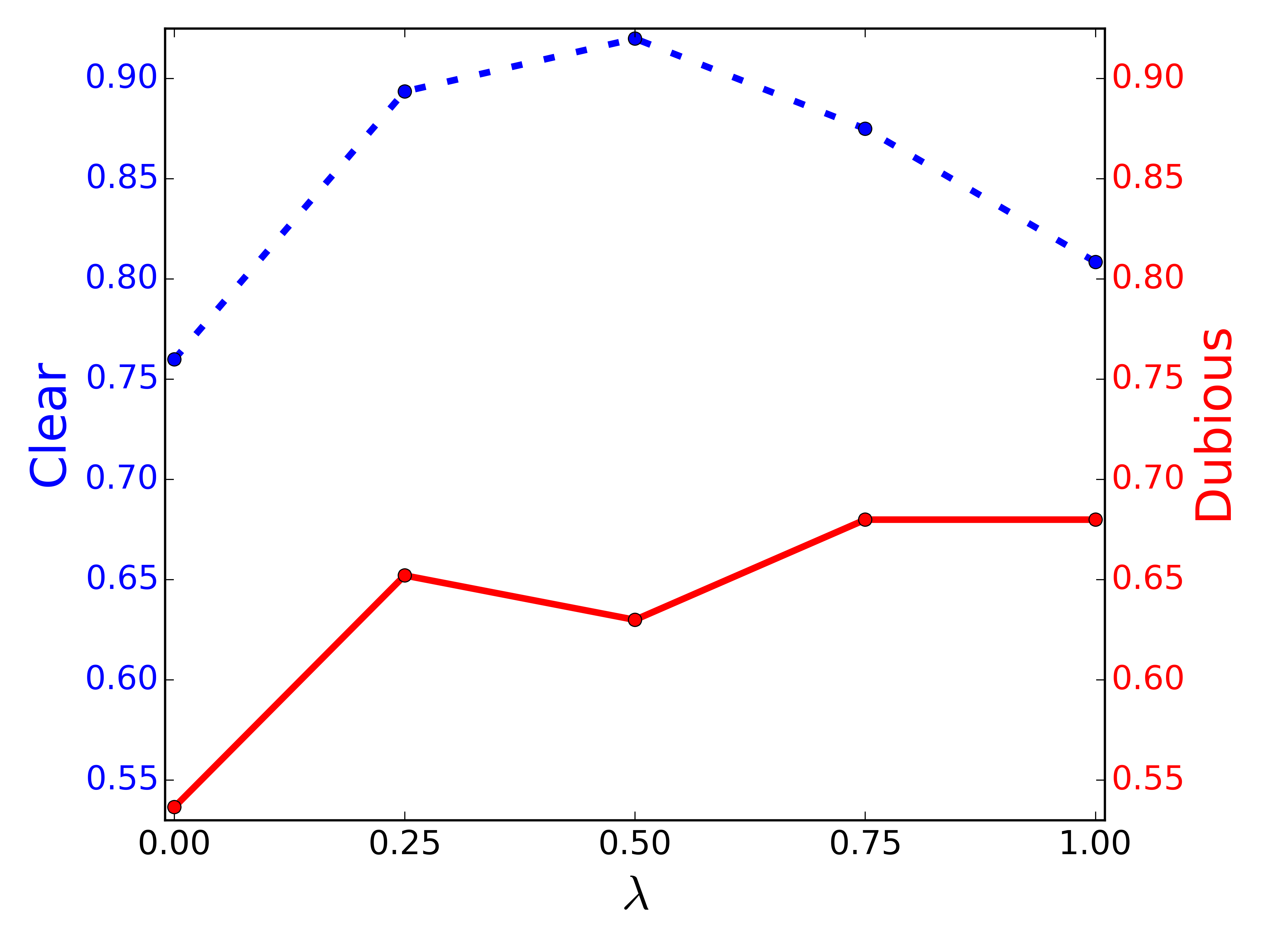} 
    \label{fig:lambda-variations}}
    \qquad
    \subfloat[]{\includegraphics[width=0.45\textwidth]{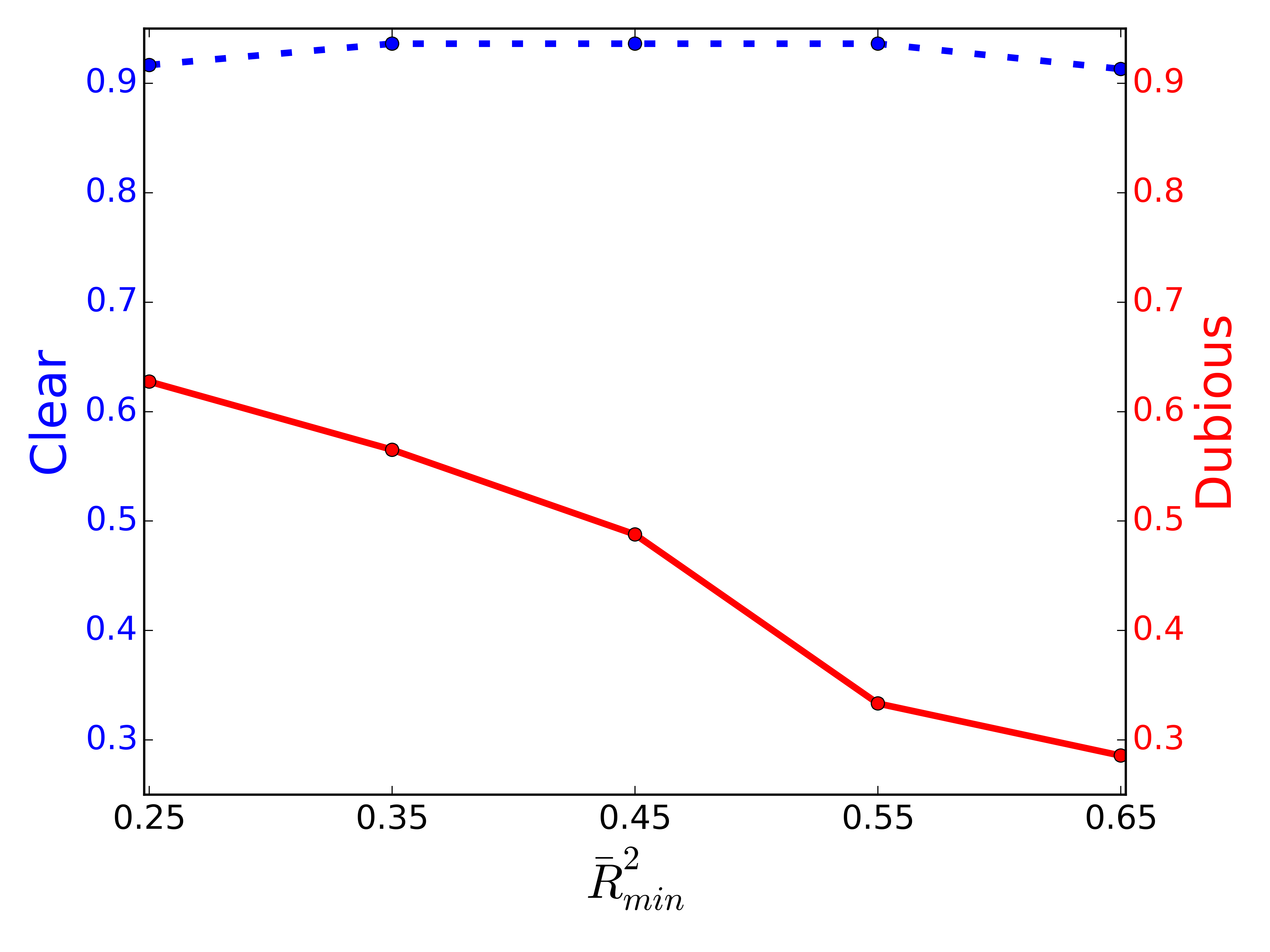} 
    \label{fig:min_adj_rsq_variation}}
\caption{F-measure values for different $\alpha$ (top left), $\beta$ (top right), $\lambda$ (bottom left), and $\bar{R}^2_{min}$ (bottom right) values. 
In all cases, values for category \textit{Clear} are in blue (dashed); for category 
\textit{Dubious}, in red (solid). \hide{It would be good to keep the vertical scale same for all plots. }}
\label{fig:coefs-var}
\end{figure}

\paragraph{Parameter $\alpha$.} To understand how robust the \textit{Outlier-Biased} scheme is, we
investigated the impact of variations in parameter $\alpha$. 
Recall that $\alpha$ controls the similarity of the weights for distinct aligned scores.  
We started with $\alpha = 0.5$ so the weights are neither too similar nor too different for distinct alignments. We then executed \tipo over the gold data set, 
and repeated the process with different $\alpha$ values, obtained with variations in the order of $10^{-2}$.
As shown in Figure~\ref{fig:robust-alpha}, variations in $\alpha$ lead
to very small changes in F-measure.  For category \textit{Clear}, the
F-measure varied less than 4\% in comparison with the initial value
($\alpha = 0.5$); for category \textit{Dubious}, less than 10\%.
This suggests that \tipo is robust with respect to slight variations in $\alpha$.

\paragraph{Parameter $\beta$.} Coefficient $\beta$ controls how many outlier errors need to be bounded 
by the model's error distribution---the higher $\beta$ is, the  
more rigorous is the discovery process. This experiment contrasts results
obtained with different choices of $\beta$. We started with the $\beta = 0.67$, executed \tipo over the gold data set, and
then repeated the process with different $\beta$ values, obtained with variations in the order of $10^{-1}$.
  The
results in Figure~\ref{fig:beta-variation} show that lower $\beta$
percentages (a less rigorous verification of meaningfulness) are more 
adequate for the gold data. To some extent, this is expected: our gold
data is derived from urban data sets, and it is likely that there are  
a number of intervening variables interfering with one another. As a 
consequence, \trend patterns may be more unclear (especially for category 
\textit{Dubious}), 
leading to higher outlier errors with respect to the original model. 
%


\paragraph{Parameter $\lambda$.} To understand the role that $\lambda$ plays in the dominant scores and 
the effectiveness of \tipo, we investigated the impact of variations 
in the order of $10^{-1}$. 
Recall that the higher $\lambda$ is, the more cumulative effects 
are taken into account.  
Results illustrated in Figure~\ref{fig:lambda-variations} show that the F-measure changes significantly 
as we vary $\lambda$. For category \textit{Clear}, average $\lambda$ values yielded the best results. As for 
category \textit{Dubious}, larger $\lambda$ values are more suitable. These results suggest that, for cases where  
\trend patterns are not so clear, as is the case with the examples in \textit{Dubious}, the use of more cumulative 
effects help discern true positives from true negatives. Our hypothesis is that matches between close values, promoted by the 
use of cumulative effects, play an important role in the detection of meaningful outlier relationships in harder scenarios. 
%

\paragraph{Parameter $\bar{R}^2_{min}$.} As discussed in Section~\ref{subsec:detection}, a model that fits its observations very poorly is 
of little use in \tipo's  verification of meaningfulness. As a consequence, we reject models whose $\bar{R}^2$ is below a certain $\bar{R}^2_{min}$. Picking an 
adequate value for $\bar{R}^2_{min}$ is difficult and the particular
value depends on the application. However, in contexts where there are many variables at play 
interfering with each other, and the goal is to \textit{find \trend patterns} instead of \textit{explaining most of the variance in the response variable},   
 moderate $\bar{R}^2$ values are expected in the models, suggesting that $\bar{R}^2_{min}$ does not necessarily need to be very high~\cite{cohen@psychbulletin1992}.  
For completeness, we decided to contrast results obtained with significantly different $\bar{R}^2_{min}$ values. We started with $\bar{R}^2_{min} = 0.25$, 
 executed \tipo over the gold data set, and then repeated the process
 with different values, obtained with variations in the order of $10^{-1}$. The results, shown in Figure~\ref{fig:min_adj_rsq_variation}, show little change for \textit{Clear}. 
In fact, the increase in $\bar{R}^2_{min}$ slightly reduced the number of false positives for this category, while keeping the number of true positives constant up until 
$\bar{R}^2_{min} = 0.55$. On the other hand, the F-measure falls drastically for \textit{Dubious} as we increase $\bar{R}^2_{min}$. In the annotation of the gold data, 
we believe that clear \trend shapes, usually associated with high $\bar{R}^2$, played an important role in the annotators' notion of meaningfulness.  
Thus, examples in which there was significant divergence have lower $\bar{R}^2$, getting more easily pruned away as $\bar{R}^2_{min}$ 
increases. 
Given that the F-measure changes very little for category \textit{Clear}, results in Figure~\ref{fig:min_adj_rsq_variation} increase the evidence that moderate   
$\bar{R}^2_{min}$ values are overall more beneficial for the type of data we examine in this paper.

\section{Related Work}
\label{related-work}

Closely related to \tipo are strategies to explain
outliers. 
%
%
MacroBase~\cite{bailis2017macrobase} classifies points in a data
stream 
as \textit{interesting} or not, and attempts to explain groups
of interesting values by highlighting correlations that most
differentiate them. 
Data X-Ray~\cite{wang@sigmod2015} identifies and explains
 errors in data that are inherent to their generation process. 
It also identifies features that best represent
erroneous elements and uses Bayesian analysis to 
 evaluate whether a set of features is associated to the cause of errors.
ExStream~\cite{zhang@edbt2017} provides explanations for 
anomalous behavior in streams detected by complex-event processing
(CEP) systems that are both concise and have high predictive value. 
%
These approaches, however, derive explanations for
\textit{categorical} features, while our goal is
to detect meaningful outlier relationships between \textit{numerical} attributes, which may help 
 explain such outliers.

Approaches have also been proposed to explain
outliers in query results.
Wu and Madden~\cite{wu2013scorpion} and Roy and
Suciu~\cite{roy@vldb2015} proposed frameworks that formulate
explanations for SQL query results as predicates over the input attributes.
Miao et al.~\cite{miao2019pvldb} proposed a system that attempts to explain outliers in one direction with related outliers in the opposite direction 
(explanations by \textit{counterbalance}). As an example, a higher than expected value may be explained by a a lower than expected value present in another 
answer tuple. 
In contrast, the explanations derived by \tipo are based on whether aligned outliers 
are 
linearly related in a statistically meaningful way. 

Also related to \tipo, albeit less directly, are attempts to
explain data anomalies without exploring data
relationships.  Dasu
et al.~\cite{dasu@kdd2014} introduced
the concept of \textit{explainable glitches}, which are similar to data
integrity violations, and proposed a non-parametric method to
empirically generate explanations for them.
Others~\cite{dang@pkdd2013,knorr@vldb1999} proposed techniques to
explain outliers in the context of a single data set, detecting which
subset of their attributes contributes most to their
\textit{outlierness}.
These ideas help formalize domain knowledge and improve the interpretability of outliers. They may be helpful in the generation of 
sophisticated outlier detection functions, which can then be combined with \tipo to provide even more complex outlier explanations. 

Techniques have also been proposed that can help explain numerical
features.  Ho et al.~\cite{ho@bigdata2016} proposed an efficient
method that, given two data sets, adaptively identifies time windows
where the data present a strong correlation. Chirigati et
al.~\cite{chirigati@sigmod2016} proposed Data Polygamy ($DP$), which
uses computational topology techniques to model spatio-temporal data
sets as scalar functions, and derives relationships for a pair of 
functions when their peaks and valleys overlap.
They focused on salient features, 
which can be common in the data --- in contrast, our goal is to find meaningful alignments that 
 can help \textit{explain outliers}.
 Note, however, that DP also detects extreme features
 (outliers) in the data and, as discussed in
 Section~\ref{subsec:dp-comparison}, it is possible to compare $DP$ 
 against \tipo. 
In fact, our experiments strongly suggest that \tipo is more suitable for the detection of important outlier relationships, 
providing insights that are compatible with users' intuitions but are filtered out by $DP$. 
Another distinction between these works and \tipo is that the latter
 makes use of cumulative effects to implicitly align extreme values
 that occur at different but close-by times, whereas the former align data based on
 an exact matching of their timestamps and thus may miss explanations
 that involve the lingering effects of extreme events.

Also related to our work are techniques for mining association rules.
Korn et al.~\cite{korn@vldb2000}~and Srikant and
Agrawal~\cite{srikant@sigmod1996} proposed techniques to mine
numerical association rules, but they did not address associations
involving numerical outliers, or verify correlations
linked to the detected rules.
There has also been work on the detection of rare 
association rules that have low support and high 
confidence~\cite{tsang2013rareassociation,yun2003rareassociation} 
but, unlike \tipo, these have focused on  associations across categorical
attributes. 

This work was motivated by issues we have encountered in urban data
with a temporal component. However, it is worth noting that while we
take timestamps into account in order to align data, our technique detects
 \trends, i.e., trends that materialize over \textit{aligned values as they grow more abnormal}, without 
 focusing on whether aligned values are temporally close. 
 Thus, approaches to find
trends and correlations for time series (see
e.g.,~\cite{hamilton1994time}) are not applicable in our scenario. In
addition, these approaches  often focus on common patterns, not on outliers. 

Finally, there are works which are related because they propose techniques similar to the ones implemented by \tipo.  
Cross-correlation functions~\cite{ccf@2006}, for example, 
are related to \tipo's attribute alignment. These functions generate alignments between data points allowing for different temporal lags, 
but no cumulative effect is taken into account. In other words, only outliers that are present in the initial data can be aligned.
In addition, cross-correlation functions are costly and the temporal lags are not  straightforward to tune. 
Rather than explicitly comparing two outliers separated by an arbitrary amount of time, \tipo calculates cumulative effects and generates new outlier 
representations. These, in turn, are aligned if they share the same timestamp. 
Heteroskedasticity~\cite{CRR73}, a common application of
\textit{WLS} that compensates for different variability in distinct
subsets of the data, is also related to a key component of \tipo: the \textit{Outlier-Biased} scheme. 
Heteroskedasticity pulls the regression towards
matching subsets with less variability which, in our context, does not necessarily 
 correspond to aligned outliers or near-outliers. 
To the best of our knowledge, our work is the first application of
\textit{WLS} that uses weights following an explicit semantics,
specific to the problem of identifying meaningful relationships 
between outliers.

\section{Conclusions}
\label{conclusions}

In this paper, we presented \tipo, a 
data-driven approach to detect meaningful outlier relationships. 
 \tipo leverages the growing number of available temporal data sets
to derive explanations for outliers by discovering meaningful relationships 
across their aligned outlies. 
To the best of our knowledge, \tipo is the first method that
attempts to explain outliers in numerical data by making use of trend
patterns (\trends).
We have developed new techniques that efficiently identify related
attributes, detect \trends across their aligned outliers, and verify
the statistical meaningfulness of these \trends. We presented an
extensive experimental evaluation, using real urban data sets, which
shows that our approach is effective in comparison with totally distinct baselines. 
Moreover, our experiments show that \tipo is robust with respect to slight parameter variations, 
and scalable.
%


There are many directions we plan to explore in future work.  While
our approach currently handles numerical attributes, it would be
interesting to investigate techniques that combine both numerical and
categorical attributes to derive more complex outlier explanations.
\tds{In this vein, we are also interested in the exploration of different regression models for the detection of \trends (e.g., random forests~\cite{murphy2013machine}), especially to capture outlier relationships 
that are nonlinear.}
We also plan on creating strategies to recommend good parametrizations for \tipo, including learning them from data. 
In addition, we would also like to examine strategies that involve the
alignment of more than two attributes, as well as the detection of nonlinear relationships across them.
We are also interested in extending \tipo for spatiotemporal data, which may be relevant for scenarios where 
data sparsity is not an issue. 
Last but not least, we plan to apply \tipo on other sources of data, increasing the diversity and depth of our case studies.

\bibliographystyle{ACM-Reference-Format}
\bibliography{paper}


\end{document}